\documentclass[12pt]{article}
\usepackage{amsmath,amssymb,latexsym,cite}
\usepackage[]{hyperref}
\usepackage{axodraw}
\input xy
\xyoption{all}

\begin{document}

\vspace*{0.5in}

\begin{center}

{\large\bf Notes on generalized global symmetries in QFT}

\vspace*{0.2in}

Eric Sharpe

Department of Physics MC 0435 \\
850 West Campus Drive\\
Virginia Tech\\
Blacksburg, VA  24061\\

{\tt ersharpe@vt.edu}
$\,$

\end{center}

It was recently argued that quantum field theories possess one-form
and higher-form
symmetries, labelled `generalized global symmetries.'  
In this paper, we describe how those higher-form symmetries
can be understood mathematically as special cases of more general
2-groups and higher
groups, and discuss examples of quantum field theories admitting actions of
more general higher groups than merely one-form and higher-form
symmetries.
We discuss analogues of topological defects for some of these higher symmetry
groups, relating some of them to ordinary topological defects.
We also discuss topological defects in cases in which the moduli `space'
(technically, a stack) admits an action of a higher symmetry group.
Finally, we outline a proposal for
how certain anomalies might potentially be understood as describing a 
transmutation of an ordinary group symmetry of the
classical theory into a 
2-group or higher group symmetry of the quantum
theory, which we link to WZW models and bosonization.

\begin{flushleft}
August 2015
\end{flushleft}

\newpage

\tableofcontents

\newpage

\section{Introduction}

The recent paper \cite{gksw} proposed a more general class of symmetries
that should be studied in quantum field theories:  in addition to 
actions of ordinary groups, it proposed that we should also consider
`groups' of gauge fields and
higher-form analogues.  For example,
Wilson lines can act as charged objects under such symmetries.
By using various defects, the paper \cite{gksw} described new 
characterizations of gauge theory phases.

Now, one can ask to what extent it is natural for $n$-forms as above
to form a group.  In particular, because of gauge symmetries, 
the group multiplication will not be associative in general, unless
perhaps one restricts to suitable equivalence classes, which does not
seem natural in general.  A more natural understanding of such
symmetries is in terms of weaker structures known
as 2-groups and higher groups, in which
associativity is weakened to hold only up to isomorphisms.

There are more 2-groups and higher groups than merely, `groups'
of gauge fields and higher-form tensor potentials (connections on
bundles and gerbes), and in this paper we will give examples of 
actions of such more general
higher groups in quantum field theory and string theory.
We will also propose an understanding of certain anomalies as
transmutations of symmetry groups of classical theories into
higher group actions on quantum theories.

To be clear, proposals for roles of 2-groups in physics have appeared
previously in a number of papers, 
in {\it e.g.} \cite{freed-higher,baez-highergauge,sati1,s3}, in
discussions of the String group (see {\it e.g.}
\cite{sp,kw,bcss,nsw}), in
the Yetter model (see {\it e.g.} \cite{yet,martins-porter}),
and in lattice gauge theories (see {\it e.g.} \cite{g-kgs1,pf1}),
to name a few examples.  Additional categorical
generalizations of (orbifold) groups, via an application of defects to
generalize the ordinary orbifold construction,
are discussed in \cite{ffrs,cr,ccr,bcp,bcp2,cv}.
See also \cite{wgw} for related ideas and applications of discrete
gauge theories and group cohomology in condensed matter physics.
The purpose of this paper is to merely to 
link the recent work \cite{gksw} to
other work on 2-groups, to review a few highlights,
and to provide a few hopefully
new results, proposals, and applications.

We begin in section~\ref{sect:overview} 
with a brief introduction to 2-groups and
higher groups, omitting mathematical details for the sake of 
readability.  In section~\ref{sect:examples:highergps} we describe
several examples of quantum field theories admitting higher group
symmetries.  We begin with a brief review of the gauge theory phase
analysis of \cite{gksw}, then describe symmetries of gauge theories with
massless matter that is invariant under a finite subgroup of the gauge
group.  (Such theories have been studied in a number of papers, and in
two dimensions naturally decompose into disjoint unions of theories,
as is reviewed in that section.)  We also review how boundary structures
in Dijkgraaf-Witten theory and also WZW models define further examples
of theories admitting higher group symmetries, albeit more complicated
higher groups than just $p$-form symmetries.  Both those examples of higher
group symmetries are discussed elsewhere in the literature, but are reviewed
here as they play a role later.  At the end of that section, we also
speculate on how current algebras in higher dimensions may define
further examples of higher group symmetries, and also mention some
other potential applications.
In section~\ref{sect:cosmo-defects} 
we briefly discuss cosmological defects in the
context of higher group symmetries, as well as `generalized moduli spaces'
(technically, stacks), defined here as `spaces' which admit actions of
higher groups.
Finally, in section~\ref{sect:anom:transmutation} we conjecture that
some anomalies might be interpreted as promoting classical ordinary-group
symmetries to quantum higher-group symmetries.

One of the ideas that drives this paper is that when the action of
an ordinary group $G$ is broken -- in mathematics, by trying to lift
$G$ to a bundle that does not admit a $G$ action, say, or in physics,
via anomalies -- there is often a larger or higher group that
will act instead.  We will encounter various manifestations of this
principle in several contexts.

An idea that has often been repeated is that one should look for 
fundamental symmetries to better understand string theory
(see for example \cite{moore-finite} for one well-known example).
The philosophy espoused by
this paper and others mentioned above is that
perhaps instead of only looking for ordinary group symmetries, we should 
also look for higher group symmetries.

\section{Overview of 2-groups and higher groups}
\label{sect:overview}

In this section, to make this paper self-contained,
we will outline definitions and examples of 2-groups
and higher groups, as these ideas are not widely known in the
physics community.  We will begin with 2-groups
(see {\it e.g.} \cite{sp,bcss,wal2,wal3,baezlauda-v,bbfw,sinh} 
for discussions and 
introductions to 2-groups).

Briefly, given objects $x$, $y$, $z$ in a 2-group, instead of demanding
that multiplication be strictly associative
\begin{displaymath}
(x \cdot y) \cdot z \: = \: x \cdot (y \cdot z),
\end{displaymath}
we instead demand that the two sides be merely isomorphic, related
by isomorphisms
\begin{displaymath}
\alpha(x,y,z): \: (x \cdot y) \cdot z \: \stackrel{\sim}{\longrightarrow} \:
x \cdot (y \cdot z) .
\end{displaymath}
Similarly, the identity axioms
\begin{displaymath}
1 \cdot x \: = \: x, \: \: \: x \cdot 1 \: = \: x
\end{displaymath}
are also replaced by isomorphisms
\begin{displaymath}
\ell_x: \: 1 \cdot x \: \stackrel{\sim}{\longrightarrow} \: x, \: \: \:
r_x: \: x \cdot 1 \: \stackrel{\sim}{\longrightarrow} \: x
\end{displaymath}
and identity relations
\begin{displaymath}
x \cdot x^{-1} \: = \: 1, \: \: \:
x^{-1} \cdot x \: = \: 1
\end{displaymath}
are also replaced by isomorphisms.  These isomorphisms must satisfy
relations of the form (see {\it e.g.} \cite{baezlauda-v})
\begin{displaymath}
\xymatrix{
& & (w \cdot x) \cdot (y \cdot z)
\ar[drr]_{\alpha(w, x, y\cdot z) }^{\sim} & & \\
( (w \cdot x) \cdot y) \cdot z \ar[urr]_{ \alpha(w\cdot x, y, z) }^{\sim}
\ar[dr]_{ \alpha(w, x, y) \otimes 1 }^{\sim} & & & &
w \cdot(x \cdot (y \cdot z)) \\
& (w \cdot (x \cdot y)) \cdot z \ar[rr]_{ \alpha(w, x \cdot y, z) }^{\sim} & &
w \cdot ( (x \cdot y) \cdot z) \ar[ur]_{1 \otimes \alpha(x,y,z) }^{\sim} &
}
\end{displaymath}
and
\begin{displaymath}
\xymatrix{
(x \cdot 1) \cdot y \ar[rr]_{\alpha(x,1,y) }^{\sim} 
\ar[dr]_{r_x \otimes 1 }^{\sim} & &
x \cdot(1 \cdot y) \ar[dl]^{1 \otimes \ell_y}_{\sim} \\
& x \cdot y & 
}
\end{displaymath}
There are technical distinctions between various descriptions of
2-groups, which will not be relevant for our purposes here (see
instead {\it e.g.} \cite{baezlauda-v}).

The description above suggests that it may be possible to construct
2-groups from a given group by weakening associativity by an automorphism
defined by an element of degree 3 group cohomology, and in fact,
this is the case, see {\it e.g.} \cite{baezlauda-v}[section 8.3],
\cite{nganter,sinh}: for a given group $G$,
group cohomology $H^3(G,U(1))$ (with trivial action on the coefficients)
defines a corresponding 2-group.  Briefly, the basic idea is that
a 3-cocycle $\alpha(g,h,k)$ defines an isomorphism
\begin{displaymath}
(g \cdot h) \cdot k \: \stackrel{\sim}{\longrightarrow} \:
g \cdot (h \cdot k)
\end{displaymath}
and as the cocycle is closed in group cohomology, the isomorphisms
so defined satisfy
the pentagon identity:
\begin{displaymath}
\alpha(g,h,k) \, \alpha(g,hk,\ell) \, \alpha(h,k,\ell) \: = \:
\alpha(gh,k,\ell) \, \alpha(g,h,k\ell) .
\end{displaymath}
We shall return to this example shortly.

Let us next discuss 2-groups appearing in discussion of
``$1$-form symmetries.''
For $G$ abelian, we can define \cite{urspriv}
the 2-groups ${\bf B}G$, ${\bf B} G_{\rm conn}$,
and ${\bf B}_{\flat} G_{\rm conn}$, which we will define
over the next several paragraphs.  
The first 2-group ${\bf B}G$ assigns\footnote{
Technically, in our conventions we will take
${\bf B} G$ to be the stack $[{\rm point}/G]$.
More generally, classifying stacks and classifying spaces naturally admit
the structure of higher groups.
We will largely avoid using the language of stacks in this paper so as to
make it more accessible, aside from technical footnotes of this form.
} to
any manifold the category of principal $G$ bundles.
There exists a tensor product that tensors any two $G$ bundles to
produce another $G$ bundle\footnote{
Technically, we should describe this in terms of tensor products
of torsors \cite{bry-bk}[section 5.1], \cite{freed-higher}, but for the 
purposes of this paper, we shall be content with looser language.
}, and the resulting operation defines a 
multiplication that is associative up to higher homotopies, as above.

The 2-group ${\bf B} G_{\rm conn}$ is the analogous classifying object 
for principal
$G$ bundles with connection, which tensors bundles and adds the
connections.  (If $G$ is finite, then as there is no nontrivial
connection on
a principal $G$-bundle, ${\bf B} G_{\rm conn} = {\bf B} G$.)
This 2-group appeared implicitly in \cite{gksw}.
Since the isomorphisms must preserve connections and so are more
restricted than the isomorphisms in ${\bf B}G$ above,
one might worry that the resulting 2-groups would all be equivalent 
to ordinary groups.  However, this is not the case.  For example,
discrete torsion implicitly provides examples of 2-groups of this
form which are not equivalent to ordinary groups.  As
discussed in {\it e.g.} \cite{medt}, discrete torsion arises from
choices of equivariant structures on $B$ fields -- one combines
the action of the orbifold group $\Gamma$
on the underlying space (here, a point) with gauge
transformations of the $B$ field.  The gauge transformations, themselves
principal $U(1)$ bundles with connection, are required
to obey the group law up to isomorphism -- in essence, a 2-group structure --
and those isomorphisms are
encoded in elements of $H^2(\Gamma,U(1))$, as connection-preserving gauge
transformations of a principal $U(1)$ bundle.  

As 2-groups, ${\bf B} G$ and ${\bf B} G_{\rm conn}$ are not
equivalent to one another \cite{nganter,urspriv}.  Now, that said,
to classify 2-groups topologically, we will sometimes replace
2-groups by their `geometric realization.'
Ultimately because
the space of connections on a fixed bundle is contractible, 
the geometric realization of ${\bf B} G$ is homotopic 
to that of
${\bf B} G_{\rm conn}$, despite the fact that the 2-groups are not
equivalent \cite{bnv}.  In both cases, the geometric realization is homotopic
to the ordinary classifying space for $G$, which we will denote as
$BG$ (with a non-boldface $B$), to distinguish it from the 2-groups above.
(See for example \cite{milnor-stasheff} for a basic discussion of the
classifying space, and appendix~\ref{app:class} for references on homotopies
of geometric realizations.)

Now, if one lifts to a loop space, some of this structure simplifies.
For example, $G$-gerbes over a space $X$ become principal $G$-bundles
on the loop space $LX$.  In particular, a Wilson line is effectively
a function on $LX$, and so obeys an ordinary group law.  Thus, if one is working
with these structures exclusively through Wilson lines, then at least
to some extent it is entirely reasonable to speak about them as ordinary
groups (involving Wilson lines) rather than higher groups, as in fact
was done in \cite{gksw}.

Finally, let us discuss ${\bf B}_{\flat} G_{\rm conn}$.
This is the analogue of ${\bf B} G_{\rm conn}$
for flat bundles.  In other words, 
${\bf B}_{\flat} G_{\rm conn}$ assigns the category of flat $G$
bundles with connection.  If $G$ is finite, then for trivial reasons
\begin{displaymath}
{\bf B}_{\flat} G_{\rm conn} \: = \: 
{\bf B} G \: = \: {\bf B} G_{\rm conn}.
\end{displaymath}

Next let us turn to further examples of 2-groups, which cannot be
understood as simply as ``$1$-form symmetries.''  In particular,
we will next consider
examples of 2-groups formed as extensions,
as described in {\it e.g.} \cite{sp}.
For $G$ a group, there are
several 2-group extensions $\tilde{G}$ given 
by\footnote{
Further discussion of the relation between these extensions can be
found in {\it e.g.} \cite{fss}.
}
\begin{displaymath}
1 \: \longrightarrow \: {\bf B} U(1) \: \longrightarrow \:
\tilde{G}_1 \: \longrightarrow \: G \: \longrightarrow \: 1 ,
\end{displaymath}
\begin{displaymath}
1 \: \longrightarrow \: {\bf B} U(1)_{\rm conn} \: \longrightarrow \:
\tilde{G}_c \: \longrightarrow \: G \: \longrightarrow \: 1 ,
\end{displaymath}
\begin{displaymath}
1 \: \longrightarrow \: BU(1) \: \longrightarrow \:
\tilde{G}_0 \: \longrightarrow \: G \: \longrightarrow \: 1 ,
\end{displaymath}
differing by whether one extends by a stack ${\bf B}U(1)$, 
${\bf B}U(1)_{\rm conn}$, or a classifying space $B U(1)$.
(One can also consider extensions by the various 
${\bf B}_{\flat} U(1)_{\rm conn}$'s,
but the examples above will suffice for this paper.)
Extensions of the form $\tilde{G}_1$ are discussed in \cite{sp}.  
In particular, this extension
can be understood as a $U(1)$-gerbe over $G$, with suitable multiplicative
structure.  For $G$ finite, these are precisely the 2-gerbes discussed
above which are 
classified by group cohomology $H^3(G,U(1))$.

Let us now suppose instead that
$G$ is simple and simply-connected; in this case,
extensions of this form above are classified by
$H^3(G,{\mathbb Z})$ (see appendix~\ref{app:class}).  
(Each type of extension is classified in the same way, as discussed
in appendix~\ref{app:class}.)
Now, for $G$ simple and simply-connected,
$H^3(G,{\mathbb Z}) = {\mathbb Z}$, and so possible extensions are
classified by an integer.  The reader may find this structure reminiscent
of WZW models, and indeed there is a close connection:  the underlying
$U(1)$-gerbe
appearing in WZW models, for which the Wess-Zumino term acts as a
coupling to the $B$ field, is precisely the 2-group $\tilde{G}_1$
above (with suitable
multiplication), and the integer classifying the extension is the same
as the level of the WZW model.  
We will return to this connection later.

So far this discussion has been rather abstract.  Let us now try to
make it more concrete by discussing more explicitly the 
multiplication on a 2-group $\tilde{G}$, built as an extension of
(simple, simply-connected) $G$ by some version of ${\bf B} U(1)$.
This product is more easily described on the loop space of $G$,
or rather a minor variation of the loop space.  (The resulting
description will also implicitly encode a version of Wilson lines for
2-groups defined as extensions $\tilde{G}$ of the form above.)

First, let us describe elements of $\tilde{G}$.
Following \cite{bry-bk}[section 6.4], we can
describe them as certain equivalence classes of pairs
$(\sigma, z)$, where
$\sigma: D^2 \rightarrow G$ for $D^2$ a two-dimensional disk
(corresponding to a filled-in loop)
and $z \in U(1)$ (corresponding to an ordinary $U(1)$ Wilson line).  
Then the product of two elements of $\tilde{G}$ (over this analogue of the
loop space) is defined by
\begin{displaymath}
(\sigma_1,z_1) \cdot (\sigma_2,z_2) \: = \:
(\sigma_1 \sigma_2, z_1 z_2 \exp\left( -2 \pi i \int_{D^2} (\sigma_1
\times \sigma_2)^* \omega\right) ) ,
\end{displaymath}
where $\omega$ is a 2-form on $G\times G$, defined as follows.
If $\nu$ is the a multiple of the
canonical 3-form on $G$ (associated with the Wess-Zumino
term in WZW models, and encoding the level), 
$p_{1,2}: G \times G \rightarrow G$ are
projection maps, and $m: G \times G \rightarrow G$ the ordinary
multiplication in $G$, then
\begin{displaymath}
p_1^* \nu \: + \: p_2^* \nu \: - \: m^* \nu \: = \: d \omega .
\end{displaymath}
(Intuitively, we can think of $\sigma$ as a semiclassical state
of the string in a WZW model, and the phase acquired in multiplication as 
the integral of the Wess-Zumino term over the path swept out by the
change in region.)
As this is defined over (a version of) the loop space,
rather than over $G$ itself,
this is an honest group structure:  associativity holds on the nose,
rather than up to cocycles.  In more detail,
associativity of the multiplication
follows from the fact that for any $\sigma_{1, 2, 3}$,
\begin{displaymath} 
\int (\sigma_2 \times \sigma_3)^* \omega \: + \:
\int (\sigma_1 \times \sigma_2 \sigma_3)^* \omega \: = \:
\int (\sigma_1 \times \sigma_2)^* \omega \: + \:
\int (\sigma_1 \sigma_2 \times \sigma_3)^* \omega ,
\end{displaymath} 
as follows from the fact that these four terms correspond schematically
to the four
faces of a tetrahedron, as illustrated schematically below:
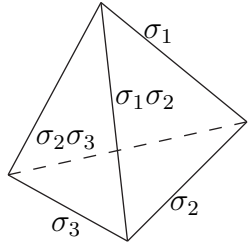
\begin{center}
\begin{picture}(100,100)(0,0)
\Line(5,30)(40,95)  \Line(40,95)(95,50)
\Line(95,50)(50,5)  \Line(50,5)(5,30)
\Line(40,95)(50,5)  \DashLine(5,30)(95,50){5}
\Text(27,7)[b]{$\sigma_3$}
\Text(72,15)[b]{$\sigma_2$}
\Text(62,80)[b]{$\sigma_1$}
\Text(57,55)[b]{$\sigma_1 \sigma_2$}
\Text(27,40)[b]{$\sigma_2 \sigma_3$}
\end{picture}
\end{center}

In passing, we should mention that there is a related story for
Lie algebroids.  Given a Lie algebroid $\overline{Q}$ associated to
some bundle $P$, the obstruction to lifting to a nontrivial
Courant algebroid $Q$ determined by an extension of $\overline{Q}$ by
$\Omega^1$:
\begin{displaymath}
0 \: \longrightarrow \: \Omega^1 \: \longrightarrow \: Q \:
\longrightarrow \: \overline{Q} \: \longrightarrow \: 0
\end{displaymath}
is given by the first Pontryagin class $p_1(P)$ \cite{bressler1}.
The relevance of this construction to the Green-Schwarz condition and
associated anomalies in heterotic strings has been discussed
in {\it e.g.} \cite{bouw,gf1,bh1}; see also \cite{sss} for a discussion of
the Green-Schwarz anomaly in present language.

In principal, $n$-groups for $n>2$ can be defined similarly to 2-groups, 
by breaking
the relations above to only hold up to higher levels of isomorphisms,
which themselves obey higher identities.  For example, to get a 3-group,
one would replace the identity
\begin{displaymath}
\alpha(w, x, y\cdot z) \circ \alpha(w \cdot x, y, z) \: = \:
(1 \otimes \alpha(x,y,z)) \circ \alpha(w,x\cdot y,z) \circ
(\alpha(w,x,y) \otimes 1)
\end{displaymath}
with isomorphisms
\begin{displaymath}
\beta(w,x,y,z): \:
\alpha(w, x, y\cdot z) \circ \alpha(w \cdot x, y, z) \:
\stackrel{\sim}{\longrightarrow} \:
(1 \otimes \alpha(x,y,z)) \circ \alpha(w,x\cdot y,z) \circ
(\alpha(w,x,y) \otimes 1)
\end{displaymath}
which themselves obey higher identities.

The theory of $n$-groups seems to be less well developed than that
of 2-groups;
we refer the reader to {\it e.g.}
\cite{nss1,nss2,urs-thesis} for a few details.
Briefly, the complications in making sense of $n$-groups are tied into the
complications of understanding higher categories, and in modern language,
are perhaps best understood by working with $\infty$-categories
(and hence $\infty$-groups).   We will not need that level of technology,
but
we will occasionally make conjectures based formally
on $n$-groups, so we will describe
a few examples we shall use in this paper, and for technical
definitions we refer interested readers to {\it e.g.} \cite{urs-thesis}.

In that spirit, for $G$ abelian,
we will define \cite{urspriv}
${\bf B}^q G$ to associate to any manifold
the $G$ $(q-1)$-gerbes on that manifold\footnote{
Readers familiar with the bar construction of classifying spaces
will recognize the deliberate
notational parallel.
}, ${\bf B}^q G_{\rm conn}$ the $G$ $(q-1)$-gerbes with
connection, 
and
${\bf B}^q_{\flat} G_{\rm conn}$ the flat $G$ $(q-1)$-gerbes with connection.
If $G$ is finite, then for trivial reasons
\begin{displaymath}
{\bf B}^q_{\flat} G_{\rm conn} \: = \: 
{\bf B}^q G \: = \: {\bf B}^q G_{\rm conn}.
\end{displaymath}

To make our notation uniform, we shall define ${\bf B}^0 G$ to be the ordinary
group of smooth maps from the manifold into $G$, 
and ${\bf B}^0_{\flat} G_{\rm conn}$ the
ordinary group of constant maps into $G$.  (The classifying space for the
former is $G$; for the latter, $K(G,0)$.)  We then recognize that in
more typical symmetry discussions in QFT, ${\bf B}^0_{\flat} G_{\rm conn}$ 
defines a
global symmetry group, and ${\bf B}^0 G$ a local symmetry group.  Analogous
parallels hold for $q>0$.

Analogues of the
extension construction above exist for higher groups 
\cite{urspriv,urs-thesis,frs}.  
For example, one can consider 
extensions of the form
\begin{displaymath}
1 \: \longrightarrow \: {\bf B}^k U(1) \: \longrightarrow \: \tilde{G}_{1,k} 
\: \longrightarrow \: G \: \longrightarrow \: 1 ,
\end{displaymath}
\begin{displaymath}
1 \: \longrightarrow \: {\bf B}^k U(1)_{\rm conn} 
\: \longrightarrow \: \tilde{G}_{c,k} 
\: \longrightarrow \: G \: \longrightarrow \: 1 ,
\end{displaymath}
\begin{displaymath}
1 \: \longrightarrow \: B^k U(1) \: \longrightarrow \: \tilde{G}_{0,k} 
\: \longrightarrow \: G \: \longrightarrow \: 1 ,
\end{displaymath}
differing by whether one extends by a higher stack
${\bf B}^k U(1)$, ${\bf B}^k U(1)_{\rm conn}$ or a space $B^k U(1)$.
As outlined in appendix~\ref{app:class}, these are classified
by $H^{k+2}(G,{\mathbb Z})$ for Lie groups $G$ of nonzero dimension, and
group cohomology
$H^{k+2}(G,U(1))$ for finite $G$.

\section{Examples of higher group symmetries in QFT}
\label{sect:examples:highergps}

In this section we list some examples of higher group symmetries
in quantum field theories.

\subsection{Review of $q$-form symmetries in gauge theories}
\label{sect:review:nati}

To make this paper self-contained,
let us begin with a very brief outline of the highlights
of some of the examples of global $q$-form symmetries (special cases of
higher group symmetries)
discussed in \cite{gksw}.

Reference \cite{gksw} gives a number of examples\footnote{
See also \cite{kovner1,kovner2,kovner3,kovner4} for related ideas.
} of global
$q$-form symmetries involving shifts of an existing gauge field.
The prototype for many of their examples was a four-dimensional $U(1)$ gauge
theory on some general four-manifold.  
This admits an (electric) action\footnote{
In more formal language \cite{urspriv}, 
we could describe the gauge field $A$ and corresponding
bundle via a map into the classifying stack ${\bf B} U(1)_{\rm conn}$.
Now, ${\bf B}U(1)_{\rm conn}$ admits an action of itself, and in particular
the substack ${\bf B}_{\flat} U(1)_{\rm conn}$, which is the action
being described above.
} of ${\bf B}_{\flat} U(1)_{\rm conn}$ 
defined by shifting the gauge field $A$ by a flat
$U(1)$ connection $A_0$, as $A \mapsto A + A_0$,
and tensoring the gauge bundle by the corresponding
flat $U(1)$ bundle.  There is also a dual magnetic action of
${\bf B}_{\flat} U(1)_{\rm conn}$, defined by
shifting the magnetic potential in the same fashion.

Note that under the action above, Wilson lines are charged, picking up
the holonomy of the element of ${\bf B}_{\flat} U(1)_{\rm conn}$ 
around the path:
\begin{displaymath}
\exp\left( \int A \cdot ds \right) \: \mapsto \:
\exp\left( \int A \cdot ds \right)
\exp\left( \int A_0 \cdot ds \right) .
\end{displaymath}
Mathematically, this can be understood in terms of passing to the
(based) loop space:
line bundles on a given space become functions on loop space,
and ${\bf B}_{\flat} U(1)_{\rm conn}$ becomes $B^0_{\flat} U(1)_{\rm conn}$, 
the group of 
constant $U(1)$ elements, on the loop space, 
so that overall this
one-form symmetry reduces to an ordinary (zero-form) group symmetry
on the loop space.  (That said, Wilson lines themselves can at least
sometimes define
a 2-group, see {\it e.g.} \cite{kap-abelcs}[section 3].)

If a Wilson line has a nonzero vev in the limit of large\footnote{
The reader might well observe that any nonzero vev of an operator that 
transforms under a symmetry defines a spontaneous
symmetry breaking.  Here, however, we intend to refer to the phase visible
in the IR, for which the limit of large loop size is relevant.  In any 
event, we follow the convention set in \cite{gksw}.
} loop sizes, 
then the one-form ${\bf B}_{\flat} U(1)_{\rm conn}$
symmetry said to be spontaneously broken.  
Conversely, if one knows that a symmetry of this form is unbroken,
one can argue that Wilson loop vevs in large loop limits must vanish,
a selection rule on Wilson lines discussed in \cite{gksw}.

Reference \cite{gksw} observes that
the gauge field $A_{\mu}$ is naturally understood as the Goldstone
boson associated with that spontaneously broken symmetry, as it undergoes
a translation under the symmetry group action.

To close the intellectual loop, it would be helpful to see explicitly
that the state
space of an abelian gauge theory has the structure of a 2-vector-space
in a representation of ${\bf B}_{\flat} U(1)_{\rm conn}$.  At a very formal
level, such a description was proposed in \cite{urs-fields07}, which we shall
outline here\footnote{
We thank U.~Schreiber for outlining his argument to us, which we repeat
here.
}.  (Readers unfamiliar with stacks may wish to
skip the rest of this paragraph.)  
Briefly, the idea is to think of states in abelian gauge
theory on a manifold $X$ (expanding about an arbitrary principal $U(1)$
bundle with connection) as defined by wavefunctions, defined on a subset
of $[X, {\bf B} U(1)_{\rm conn}]$ satisfying a polarization condition.
Using ``crit'' to label the pertinent subset, the wavefunctions are elements
of the 2-vector-space
\begin{displaymath}
\left[ [X, {\bf B} U(1)_{\rm conn}]_{\rm crit}, {\mathbb C} \right] ,
\end{displaymath}
which naturally
admits an action of ${\bf B}_{\flat} U(1)_{\rm conn}$ \cite{urspriv}. 
More globally, we are under the general impression that the representation
theory of higher groups is still under development, so in the rest of this
paper we will occasionally make physics-based conjectures regarding
representations, but for the most part we will not be able to make
strong statements of results concerning representations.

Some related examples discussed in \cite{gksw} include:
\begin{itemize}
\item A four-dimensional $G$ gauge theory admits a global action of
${\bf B}_{\flat} Z(G)_{\rm conn}$ defined by shifting the gauge field by a 
flat\footnote{
$Z(G)$ denotes the center of $G$.
}
$Z(G)$ connection, and similarly tensoring the gauge bundle by the
corresponding flat $Z(G)$-bundle.
\item A theory with a dynamical $q$-form potential
admits a global action
of ${\bf B}^q_{\flat} U(1)_{\rm conn}$ given by shifting 
the $q$-form potential by a flat
$q$-form, and tensoring the $q$-gerbe by the corresponding flat
$q$-gerbe.  (See also \cite{bgru} and references therein for related examples.)
\end{itemize}
In all these cases, various defects\footnote{
See for example \cite{taffy} for a discussion of defects in topological
field theories.
} play a role analogous to charged
particles for ordinary symmetries.

In any event, the paper \cite{gksw} focused on analyzing gauge theory
phases using the ideas above, whereas in this paper we will look in
different directions.

\subsection{Gauge theories with subgroup-invariant
massless matter}
\label{sect:gerbes}

In this section we shall study gauge theories in which a finite subgroup
of the center of the gauge group acts trivially on massless matter,
and related orbifolds,
as further examples of theories with generalized global symmetries.

As a warm-up, consider the two-dimensional
supersymmetric ${\mathbb C}{\mathbb P}^n$ model, consisting of a $U(1)$
gauge theory with $n+1$ charged massless chiral multiplets, 
but rather than giving
all the chiral multiplets charge $1$ as in the usual construction, instead
give them
all charge $k$.  
Perturbatively, this would appear to be the same as the ordinary
${\mathbb C}{\mathbb P}^n$ model, in which all matter fields have
charge $1$, but in fact these cases can be distinguished:
\begin{itemize}
\item One way to distinguish these theories is by adding massive
fields of charge $\pm 1$ to the theory with massless charge $k$ matter.
The existence of this massive matter can still be detected even below the cutoff
scale, via
the periodicity of the theta angle, which acts as an electric field in
two dimensions.  To see this, 
simply build a capacitor in the two-dimensional theory; as one
increases its size, one can excite arbitarily massive matter, and so
the theta angle periodicity provides a test for charged matter beyond the
renormalization cutoff scale.  (For more information, see for example
\cite{glsm,nr,msx}, where this model was originally discussed.)
\item In principle, one could use defects to distinguish these theories.
For example, one could specify that the theory contains Wilson lines
in the charge $1$ representation.  Such Wilson lines would not be
well-defined after dividing the charges by $k$, and so specifying their
existence would distinguish these theories.  Such structures were
discussed in four-dimensional theories
in the context of discrete theta angles in \cite{ast,gmn}.
\item In principle, the global structure of the gauge group could also
be detected via gauge transformations on non-simply-connected
spacetimes.  Put another way, if the spacetime is topologically nontrivial,
then to specify matter fields, one must specify a bundle to which
the matter couples, not just a representation.  The specification of the
bundle eliminates any ambiguities, and as different bundles lead to
different zero modes and different anomalies, one also sees distinct
physics.  This is also discussed in for example \cite{glsm,nr,msx}.
\end{itemize}  

At low energies, below the scale of any massive minimally charged matter,
this theory has a ${\bf B} {\mathbb Z}_k$ symmetry, 
acting by translations of the
$U(1)$ gauge field in a ${\mathbb Z}_k$ subgroup, as this leaves the
action invariant.  (This same group may act nontrivially on defects that
distinguish the charge $k$ and charge $1$ massless matter cases; however,
a nontrivial action on such operator vevs should be interpreted in terms
of spontaneous symmetry breaking, rather than explicit symmetry
breaking.)  This ${\bf B} {\mathbb Z}_k$ action is of the same form
as discussed in the last section:  the gauge bundle is `tensored' with
a ${\mathbb Z}_k$ bundle, for example.  However, at higher
energies, this ${\bf B} {\mathbb Z}_k$ symmetry is explicitly broken by any 
massive minimally charged matter, as the action is no longer invariant.

We could also discuss the theory without massive minimally charged matter
or defects fixing the global structure of the gauge group.
This theory, in which the charges of all matter fields are multiples of $k$,
naively appears to also have a ${\bf B} {\mathbb Z}_k$ symmetry, 
but this symmetry
is an artifact of the charge scaling, and so has no physical significance.

Another set of examples involves orbifolds by finite groups that do not
act effectively on the target space.  A simple example, discussed
extensively in {\it e.g.} \cite{nr}, is the orbifold $[X/D_4]$,
where $D_4$ is the eight-element group described as an extension
\begin{displaymath}
1 \: \longrightarrow \: {\mathbb Z}_2 \: \longrightarrow \: D_4 \:
\longrightarrow \: {\mathbb Z}_2 \times {\mathbb Z}_2 \: 
\longrightarrow \: 0,
\end{displaymath}
and where $D_4$ acts on $X$ by first projecting to ${\mathbb Z}_2 \times
{\mathbb Z}_2$, so that the remaining ${\mathbb Z}_2$ acts trivially
on $X$.  By computing partition functions, it is simple to demonstrate
that this theory is not the same as the $[X/{\mathbb Z}_2 \times
{\mathbb Z}_2]$ orbifold -- both theories admit modular-invariant
partition functions, but those modular-invariant partition functions
are different.  In particular, if $H$ is a subgroup of the orbifold group
that acts completely trivially on the space, then the theory admits
a ${\bf B}H$ action.  In this case, the $[X/D_4]$ orbifold admits a
${\bf B}{\mathbb Z}_2$ action.

It is straightforward to see the ${\bf B} {\mathbb Z}_k$ actions in orbifold
examples, but for completeness, we shall work through two examples here.
\begin{itemize}
\item First, consider an orbifold $[X/{\mathbb Z}_k]$ where all of the
${\mathbb Z}_k$ acts trivially on $X$.  In other words, for all
$x \in X$ and all $g \in {\mathbb Z}_k$,
$g \cdot x = x$.  The one-loop partition function
of this theory is involves a sum over principal ${\mathbb Z}_k$
bundles, of the form
\begin{displaymath}
Z \: = \: \frac{1}{| {\mathbb Z}_k | } 
\sum_{g,h \in {\mathbb Z}_k} Z(X) \: = \: | {\mathbb Z}_k | Z(X).
\end{displaymath}
(A principal ${\mathbb Z}_k$ bundle on $T^2$ is specified by a commuting
pair of elements of ${\mathbb Z}_k$, hence the sum over $g, h
\in {\mathbb Z}_k$.)
The action of ${\bf B}{\mathbb Z}_k$ is merely to rotate 
the bundles amongst themselves.  For example, given another bundle
defined by a (commuting) pair $(g',h')$, the action on the bundle
defined by the pair $(g,h)$ is merely 
\begin{displaymath}
(g,h) \: \mapsto \: (g' g, h' h).
\end{displaymath}
\item A somewhat less trivial example is provided by the
$[X/D_4]$ example given above.  This theory admits a ${\bf B}{\mathbb Z}_2$
action, which we can see in the one-loop partition function as follows.
Write the one-loop partition function as
\begin{displaymath}
Z \: = \: \frac{1}{| D_4 |} \sum_{g,h \in D_4, \: \: gh = hg}
Z_{g,h}(X) ,
\end{displaymath}
where the sum is over commuting pairs of elements of $D_4$ ({\it i.e.}
principal $D_4$ bundles on $T^2$), and $Z_{g,h}(X)$ denotes the partition
function with boundary conditions determined by $g$, $h$.
Now, a ${\mathbb Z}_2$ bundle is determined by a pair of elements of
${\mathbb Z}_2 \subset D_4$, whose elements we will denote $\{1, z \}$,
where $z^2 = 1$.
Thus, a general ${\mathbb Z}_2$ bundle is given by a pair $(z^m,z^n)$
for integer $m$, $n$.  Such a bundle will act on a $D_4$ bundle $(g,h)$
as
\begin{displaymath}
(g,h) \: \mapsto \: (g z^m, h z^n ) .
\end{displaymath}
Since the ${\mathbb Z}_2$ acts trivially on $X$,
\begin{displaymath}
Z_{g z^m, h z^n} \: = \: Z_{g,h} ,
\end{displaymath}
and so the partition function is preserved.
\end{itemize}
More generally, for any $G$ orbifold of a space $X$, if a subgroup
$H$ acts trivially on $X$, then in the same fashion as above,
${\bf B}H$ defines a symmetry of the theory, as can be seen in rotations
of the $G$ bundles by $H$ subbundles.

Four-dimensional analogues were discussed in \cite{hs}.  Two examples of this
form are as follows:
\begin{itemize}
\item A $U(1)$ gauge theory with $N$ matter fields of charge $+k$ and
$N$ matter fields of charge $-k$.
\item An $SU(2)$ gauge theory with adjoints.
\end{itemize}
At low energies, generically along the Higgs branch, the second reduces
to a $U(1)$ gauge theory with matter of charge divisible by $2$,
{\it i.e.} an example of the former theory.
As in two dimensions, these four-dimensional theories can be distinguished
from minimally-charged matter theories in several ways:
\begin{itemize}
\item One option is to add massive minimally-charged matter to the theory,
as in two dimensions.  The presence of such matter can no longer be
sensed by the theta angle periodicity, as the theta angle no longer
acts as an electric field, but instead in a theory coupled to gravity,
one can use Reissner-Nordstrom black holes to similar effect, as
discussed in \cite{hs}.
\item Another option is to specify a set of defects in the theory which
are well-defined only for certain global gauge groups.  This was the
strategy followed in \cite{ast,gmn} to distinguish $SU(2)$ from $SO(3)$
theories, for example.
\item Finally, as before, if the spacetime is topologically nontrivial,
then a unique specification of the matter will have the same ffect.
\end{itemize}

In two dimensions, gauge theories in which a subgroup of the gauge
group acts trivially on massless matter are equivalent to theories
with restrictions on nonperturbative sectors.
In four dimensions, the analogue is more complicated.
This matter is discussed
in greater detail in \cite{hs,decomp}.  These theories can also be understood,
in different language still, in terms of QFT's coupled to TFT's, though we
shall not use that language here.

There are many other related examples, in which massless matter
is invariant under a subgroup $G$ of the center of the gauge group,
and these examples all have a ${\bf B}G$ symmetry at low energies.

These theories are sometimes known as gerbe theories, because they are
typically sigma models on gerbes as discussed in, for
example, \cite{msx,hhpsa}.  Very briefly, for those readers who are
curious, a sigma model on a (Deligne-Mumford) stack 
is defined by first picking a presentation
of the stack as a global quotient $[X/G]$ of some ordinary space $X$ by some
group $G$, which need not be finite and need not act effectively
(but whose stabilizers on $X$ are finite).
The `sigma model on the stack' is then a $G$-gauged sigma model on $X$,
or rather, its universality class.  If a subgroup $H$ of $G$ acts trivially
on $X$, then the stack is known technically as an $H$ gerbe, and the theory is
a sigma model on a gerbe.

We have already seen how theories of this form admit ${\bf B}H$ actions --
by rotating principal $G$ bundles by $H$ subbundles, which (as $H$
acts trivially) leaves the theory invariant.  (This is true for both
gauge theories and orbifolds, and we have seen examples of each.)
As sigma models on gerbes,
we can view these ${\bf B}H$ structures 
another way.  An $H$-gerbe (for $H$ finite) over a space is, mathematically,
the total
space of a ${\bf B}H$ bundle.  As such, ${\bf B}H$ acts on the fibers of
the bundle, and so acts on the gerbe and, physically, on the sigma model.

The same language gives another perspective on sigma models on gerbes.
As has been discussed elsewhere (see {\it e.g.} \cite{hs}), in two dimensions
these are equivalent to sigma models with restrictions on nonperturbative
sectors.  One way to think about the origin of those restrictions is
as a requirement that the sigma model maps preserve the 
${\bf B} H$ invariances.

The Higgs moduli spaces of these theories also have gerbe structures, and so
admit actions of higher groups.  We will explore this point in greater
detail in section~\ref{sect:genl-moduli-spaces}, where such moduli
spaces will be described as `generalized moduli spaces.'

So far we have discussed some basic examples.  It has been shown that
dualities break the
${\bf B}G$ symmetries discussed in this section.  
Specifically, 
it has now been established that these two-dimensional
theories 'decompose' (via a form
of T-duality) into
disjoint unions of simpler theories, without these ${\bf B}G$ symmetries,
which solves a technical issue with cluster decomposition,
see {\it e.g.} \cite{hhpsa} for a discussion of decomposition in
two-dimensional nonlinear sigma models, \cite{decomp} for a discussion of
decomposition in two-dimensional nonabelian gauge theories,
and \cite{hetstx} for
a heterotic version.  This decomposition makes predictions for
Gromov-Witten invariants, which have been checked rigorously
(see {\it e.g.} \cite{ajt1,ajt2,ajt3,ajt4,gt1,tseng1}), 
and also plays a role in understanding
phases of certain GLSMs \cite{hhpsa,cdhps,hori2,hkm} 
(see \cite{es-sm14} for a more
complete list of references and reviews).   
It would be interesting to understand if there are analogues of decomposition
for
any notion of ${\bf B}G$ gauge theories in some dimension.
In any event, reference~\cite{hs} contains a 
study of how these gerbe structures over moduli spaces vary under
dualities.

\subsection{Boundary structures in Dijkgraaf-Witten theory}
\label{sect:boundary-dw}

Recall that for finite $G$, an element of group cohomology
$H^3(G,U(1))$ (with trivial action on the coefficients) defines
a 2-group $\tilde{G}$ as the extension
\begin{displaymath}
1 \: \longrightarrow \: {\bf B}U(1) \: \longrightarrow \:
\tilde{G} \: \longrightarrow \: G \: \longrightarrow \: 1 .
\end{displaymath}
In this section, we shall review arguments that this 2-group $\tilde{G}$ acts
on the boundaries in three-dimensional
Dijkgraaf-Witten theories \cite{dw} defined by the
corresponding element of $H^3(G,U(1))$.  (We emphasize that the
observations of this subsection are not original to us, but instead are
reviewed for completeness and later use in this paper.  
See instead {\it e.g.} \cite{urs-thesis}[p. 757] for references.)   

First, consider the special case of a 
trivial $G$ gauge theory in two dimensions, for $G$ finite
 -- a $G$-orbifold -- with
discrete torsion.  Discrete torsion is defined by an element of 
$H^2(G,U(1))$ (with trivial action on the coefficients), and elements of
that
same group cohomology group determine a central extension $\tilde{G}$:
\begin{displaymath}
1 \: \longrightarrow \: U(1) \: \longrightarrow \: \tilde{G} \: 
\longrightarrow \: G \: \longrightarrow \: 1 .
\end{displaymath}
As was
argued in \cite{dougdt1,dougdt2,medt,pauldt}, the extension $\tilde{G}$
acts on the Chan-Paton factors and D-branes.  The extension
$\tilde{G}$ is still an ordinary group, but this example
provides a prototype for the 2-group action that will appear
on boundaries of three-dimensional Dijkgraaf-Witten theories.

Abstractly, to define
the two-dimensional $G$ gauge theory in the presence of the boundary,
one must express how the group $G$ acts on the Chan-Paton factors as
encoded in a Wilson line along the boundary (or equivalently a bundle
$L$ with connection).  If the $G$ action
lifts honestly, meaning that there are isomorphisms
$\psi_g: g^* L \rightarrow L$ such that the $\psi_g$'s obey the
group law, then $L$ is said to admit a $G$-equivariant
structure.  In general, however, the group law might not be obeyed,
meaning the lift of $G$ to the boundary theory is obstructed.
Such obstructions are encoded in elements\footnote{
In this context, $H^2(G,U(1))$ is known as Mumford's group of $L$ or
the theta group of $L$ \cite{tonypriv}.
} of $H^2(G,U(1))$.
In such a case, although the action of $G$ on the boundary theory
is obstructed, there is an action of $\tilde{G}$ (the lift associated
with the same element of $H^2(G,U(1))$) on the boundary theory:
the boundary theory is always $\tilde{G}$-equivariant, for the extension
determined by the obstruction to $G$-equivariance.

There is a closely analogous story in three-dimensional
Dijkgraaf-Witten theories \cite{dw}.
To help clarify the discussion, we shall assume that the three-dimensional
finite $G$ gauge theory includes a trivial action on a nonlinear sigma
model on some space $M$.  (It will become clear that $M$ is irrelevant.)
Then, a boundary in the theory should have
an action which includes a term of the form
\begin{displaymath}
\int \phi^* B
\end{displaymath}
for some two-form potential $B$, and map $\phi$ from the three-dimensional
spacetime into $M$.  Assuming that the boundary of the
boundary is empty, this theory has a natural action of the 2-group
${\bf B}U(1)_{\rm conn}$, 
which acts on $B$ by gauge transformations of the familiar form
$B \mapsto B + d \Lambda$.
However, this is only part of the 2-group that acts on the boundary
of this theory.

Since the bulk contains a $G$ gauge theory, to fully describe the theory,
we must describe how $G$ acts on the boundary.  In general, $G$ can
be combined with gauge transformations of the $B$ field,
{\it i.e.} elements of ${\bf B}U(1)_{\rm conn}$.  
As discussed elsewhere, however,
associativity in ${\bf B}U(1)_{\rm conn}$ holds only up to isomorphism.  If the
isomorphisms cannot be trivialized, then we do not have any sort of
homomorphism $G \rightarrow {\bf B}U(1)_{\rm conn}$, 
but rather a more complicated
structure comes into play.  

\begin{figure}[h]
\begin{center}
\begin{picture}(100,100)(0,0)
\Line(5,30)(40,95)  \ArrowLine(40,95)(95,50)
\ArrowLine(95,50)(50,5)  \ArrowLine(50,5)(5,30)
\Line(40,95)(50,5)  \DashLine(5,30)(95,50){5}
\Text(27,7)[b]{$g$}
\Text(72,15)[b]{$h$}
\Text(62,80)[b]{$k$}
\end{picture}
\end{center}
\caption{A 3-simplex.  \label{fig:3-simplex}}
\end{figure}
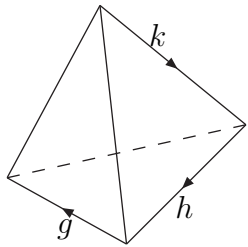

We can link the breakdown in associativity in those boundary gauge
transformations to the group cohomology defining the bulk action.
Recall from \cite{dw}[section 6.5] that the action associated to a 3-simplex
such as that in figure~\ref{fig:3-simplex}
is given by a group cohomology 3-cocycle $\alpha(g,h,k)$.  
In fact, by imagining that the 3-simplex contains a boundary just as in
\cite{pauldt}, 
we can identify the boundary phase with $\alpha(g,h,k)$.

The total action in the case that spacetime is a three-torus, for example,
can be computed from a simplicial decomposition of $T^3$, and for
fixed group elements on the boundaries is given
in \cite{dw}[equ'n (6.35)] as
\begin{displaymath}
\frac{
\alpha(g,h,k) \alpha(h,k,g) \alpha(k,g,h) 
}{
\alpha(g,k,h) \alpha(h,g,k) \alpha(k,h,g)
} .
\end{displaymath}
This is the same phase factor that was computed for formal 
$C$-form analogues of discrete torsion in \cite{cdt}.
It is invariant under group coboundaries, as well as $SL(3,{\mathbb Z})$.
The interpretation of the 3-cocycle $\alpha(g,h,k)$ can be read off
from\footnote{
Although \cite{cdt} was written to describe group actions on $C$ fields,
with only minor modifications it applies to the present case, essentially
by identifying the $B$ fields for various $g$ as the image of the boundaries,
rather than as gauge transformations themselves.
} \cite{cdt}:  it is precisely the obstruction to associativity in
the ${\bf B}U(1)_{\rm conn}$ transformations.  
Finally, note that although the nonlinear
sigma model $M$ played a role in initially setting up our discussion,
it is irrelevant to the result.

Putting this together, we see that in Dijkgraaf-Witten theory on a manifold
with boundary, the gauge group $G$ can not act honestly on the boundary
theory, but rather associativity of its group law is obstructed by an
element of group cohomology $H^3(G,U(1))$, the same element defining the
Dijkgraaf-Witten action.  

Now, although $G$ itself can not act honestly on the boundary theory,
the element of $H^3(G,U(1))$ defines a 2-group extension
\begin{displaymath}
1 \: \longrightarrow \: {\bf B}U(1) \: \longrightarrow \:
\tilde{G} \: \longrightarrow \: G
\: \longrightarrow \: 1 ,
\end{displaymath}
which does act on the boundary theory -- the obstruction to 
associativity indicates that one should replace $G$ by $\tilde{G}$
above.

It is worth mentioning that
a closely related argument for ordinary discrete torsion in 
two dimensions appears
in \cite{pauldt}.  
There, the fact that the group 2-cocycle defines a boundary phase factor
was applied to derive an expression for a 2-simplex
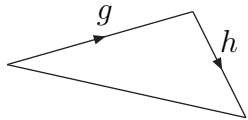
\begin{center}
\begin{picture}(100,50)(0,0)
\ArrowLine(5,25)(75,45)  \ArrowLine(75,45)(95,5)
\Line(95,5)(5,25)
\Text(42,40)[b]{$g$}  \Text(89,30)[b]{$h$}
\end{picture}
\end{center}
which was then applied to compute phase factors for Riemann surfaces
of various genera, obtaining results matching standard discrete torsion
phase factors.  The conclusion there was analogous:  the same element
of $H^2(G,U(1))$ that defines discrete torsion, also defines an
extension of $G$ that acts honestly on the boundary theory.
In that case, the extension is an ordinary group extension, whereas
here the extension that acts on the boundary theory, is a 2-group extension.

The same argument appears to imply that boundaries of 
higher-dimensional analogues
of Dijkgraaf-Witten theory, defined by elements of $H^d(G,U(1))$
in $d$ bulk dimensions,
should be acted upon by $(d-1)$-groups given as extensions
\begin{displaymath}
1 \: \longrightarrow \: {\bf B}^{d-2} U(1) \: 
\longrightarrow \: \tilde{G}
\: \longrightarrow \: G \: \longrightarrow \: 1 ,
\end{displaymath}
Indeed, precisely this structure has been discussed elsewhere, see
for example \cite{urs-thesis}[section 3.9.130].

A different approach to Dijkgraaf-Witten theory and 2-groups is
discussed in \cite{martins-porter}.  The approach there seems to be
to extend Dijkgraaf-Witten theory to 2-groups, rather than 
consider boundaries in Dijkgraaf-Witten theory, as we have reviewed here.
A related recent paper \cite{monnier} describes $B^p G$ gauge theories,
for $G$ finite.

\subsection{WZW models}
\label{sect:ex:wzw}

As seems to be very well-known in certain circles
(see {\it e.g.} \cite{bcss,urs-thesis}),
a different example of a theory with a type of one-form (2-group) action is
provided by WZW models \cite{ed-wzw}.  For applications later in this paper,
we shall give a physics-oriented review
of the highlights here.
Recall that a WZW model in two
dimensions is essentially a sigma model on a group manifold $G$ with
a $B$ field with nonzero curvature, where the $B$ field is encoded in
the Wess-Zumino term via its curvature as
\begin{displaymath}
\int_{\Sigma} g^* B \: = \: \int_Y {\rm Tr}\, (g^{-1} dg)^3
\end{displaymath}
for a three-manifold $Y$ bounding $\Sigma$.
(Strictly speaking, the Wess-Zumino term is defined up to addition of
a closed\footnote{
In a nonlinear sigma model on a space $X$, a closed $B$ field defines
an analogue of a $\theta$ angle, twisting nonperturbative sectors by
phases given by the holonomy of the $B$ field on various 2-cycles,
which is trivial only when $B$ is exact.  Here, however, we assume
$G$ is simple and simply-connected, in which case both $H^2(G,{\mathbb Z})$
and $H_2(G,{\mathbb Z})$ vanish.   
}
2-form; locally,
\begin{displaymath}
\int_{\Sigma} g^* B \: \sim \: \int_{\Sigma} g^* (B \: + \: d \Lambda)
\end{displaymath}
for some one-form $\Lambda$.)
We claim that a WZW model on $G$ admits an action of the 2-group\footnote{
Because we want to preserve the $B$ field defined intrinsically by the
Wess-Zumino term, we only consider 2-groups defined by extensions by
${\bf B} U(1)$, and not ${\bf B} U(1)_{\rm conn}$.
} 
\begin{displaymath}
1 \: \longrightarrow \: {\bf B}U(1) \: \longrightarrow \: 
\tilde{G}_1 
\: \longrightarrow \: G \: \longrightarrow \: 1 ,
\end{displaymath}
where the choice of extension of $G$ by ${\bf B} U(1)$
is the level of the WZW model.  In this case, the stack underlying the
2-group is the $U(1)$ gerbe whose connection $B$ is defined
above.

As a quick consistency check, note that such extensions $\tilde{G}_1$
are classified by (appendix~\ref{app:class}) 
elements of $H^3(G)$, which also classify the
allowed levels of the WZW model (for simple simply-connected $G$,
which we assume).

This 2-group is merely acting by automorphisms of the underlying gerbe.
In this case, the 2-group is not merely
``$G$-valued one-forms,'' but rather is more complicated.

Physically\footnote{
Mathematically,
the 2-group action on the gerbe is described in \cite{bry-bk}[section 7.3],
and its multiplication in terms of Wilson lines ({\it i.e.} on the loop space)
was described explicitly in section~\ref{sect:overview}.
}, we can describe 
the action of $\tilde{G}_1$ by right- or left-
multiplication on the fields of the WZW model as follows.
First, describe a multiplication on $\tilde{G}_1$, {\it i.e.} a map
$\tilde{G}_1 \times \tilde{G}_1 \rightarrow \tilde{G}_1$, 
by a pair $(h, A)$ consisting
of a group element $h \in G$ and a connection $A$ on a principal $U(1)$
bundle over $G$.  (The effect of multiplication by an element of
$\tilde{G}_1$ is to induce an automorphism of $\tilde{G}_1$, and such 
automorphisms are essentially specified by pairs of the given form.)
To describe the action on the WZW model, in
closed string path integral quantization, we must specify the action on
the classical map $g: \Sigma \rightarrow G$ ($\Sigma$ the worldsheet)
and on the WZW model $B$ field.  These actions are as follows:
\begin{itemize}
\item $g \mapsto hg$ or $gh$ (depending upon whether a left- or 
right-multiplication was specified),
\item $B \mapsto B + F$, where $F$ is the curvature of $A$.
\end{itemize}
The action on $g$ itself is identical to the $G$ action in an ordinary
WZW model, and has been extensively discussed elsewhere.
The action on $B$ leaves the curvature of $B$ invariant, and hence the
Wess-Zumino term is unaffected.  In principle, the $B$ field periods have
changed by $\int g^* F$, but since $F$ lies in integral cohomology,
and the $B$ field curvature is multiplied by an integer (the level),
the closed string theory is invariant.

Put more simply, the action of $\tilde{G}_1$ on the fields
of the WZW model is nearly identical to the action of $G$ itself, the 
difference being that the former includes an additional $B$ field
transformation.  The reason such a similar structure is possible is 
mathematically that the gerbe admits a canonical
$G$-equivariant structure, and so $G$ has a natural action on the gerbe
defining the WZW model.  That said, the $B$ field transformations
induce subtle differences in phases which manifest elsewhere.
For example, if we apply the same reasoning to states
defined on hemispheres, then the $B$ field transformations generate
phase rotations on states, which the $G$ action by itself would not provide.
Thus, in principle states in canonical quantization detect the difference
between the actions of $G$ and $\tilde{G}_1$, via differences in phases. 
(Also, the $B$ field actions of the $\tilde{G}_1$ multiplication also
manifest in D-brane actionas, as has been discussed in
\cite{fss2}.)

Now, 
left- and right-multiplication by elements of $G$ is a global symmetry
(though not a local symmetry) of the classical theory.  
However, in the WZW model, the algebra of their left- and
right-moving Noether currents $J^a$ has
a central extension, which in a dual theory of fermions would correspond to
a two-point anomaly.
As is well-known, 
the resulting symmetry algebra is a Kac-Moody
algebra, of the form
\begin{equation}   \label{eq:kac-moody}
\left[ J_n^a, J_m^b \right] \: = \:
f^{abc} J^c_{n+m} \: + \: \frac{1}{2} k n \delta^{ab} \delta_{n+m,0} ,
\end{equation}
where the $f^{abc}$ are the structure constants of the (finite-dimensional)
Lie algebra.  As a result of the central extension, the currents $J^a$
do not transform in a representation of the group $G$, unlike primary
fields, but rather
transform under the group action as \cite{gw}[equ'n (3.17)]
\begin{displaymath}
\delta_{\omega} J(z) \: = \: [ \omega(z), J(z)] \: + \: \frac{1}{2} k \omega'(z)
\end{displaymath}
for $\omega$ a Lie-algebra-valued function.

Those Kac-Moody algebras have a natural interpretation in terms of
the 2-group $\tilde{G}_1$, as infinitesimal automorphisms, as has
been discussed in {\it e.g.} \cite{bcss}.  Now, Kac-Moody algebras are perhaps
more widely understood in terms of the algebra of an extension of
the algebra of the loop group $LG$ of $G$, but these are closely
related.  Specifically, the looping of the 2-group $\tilde{G}_1$,
\begin{displaymath}
1 \: \longrightarrow \: {\bf B} U(1) \: 
\longrightarrow \: \tilde{G}_1 \:
\longrightarrow \: G \: \longrightarrow \: 1
\end{displaymath}
is precisely\footnote{
See \cite{bcss,wal2,wal3} for further discussion of the relationship between
loop groups and 2-groups.  It should be noted that for this discussion,
we specifically need $\tilde{G}_1$ and not $\tilde{G}_c$, as the latter is
not so closely related to the loop group extension.}
the analogous central extension $\widetilde{LG}$ of the loop group $LG$,
\begin{displaymath}
1 \: \longrightarrow \: U(1) \: \longrightarrow \: \widetilde{LG} \:
\longrightarrow \: LG \: \longrightarrow \: 1.  
\end{displaymath}
(This is a special case of the statement that a $U(1)$ 1-gerbe over a space $X$
lifts to a principal $U(1)$ bundle over the loop space $LX$.)
In this fashion, we can understand the more historically common interpretation
of the Kac-Moody algebra, in terms of extensions of $LG$, as a looping
of the 2-group $\tilde{G}_1$. 
Phrased differently, the de-looping of $\widetilde{LG}$, though not
a group, is the 2-group $\tilde{G}_1$.

So far we have briefly reviewed how the 2-group $\tilde{G}_1$ has a natural
action on the WZW model by combining the $G$ action with $B$ field
gauge transformations (in fact, the underlying $U(1)$ gerbe is the
one that defines the Wess-Zumino term), and that the Kac-Moody algebra
is the infinitesimal algebra of that same 2-group.

The reader might ask at this point about the Noether current associated
with this 2-group $\tilde{G}_1$ and its left- and right-multiplications.  
Since the action of $\tilde{G}_1$ on the fields of the WZW model is, in
the closed string theory, effectively indistinguishable from the action of
$G$, the Noether currents for $\tilde{G}_1$ should be the same as that
for $G$ -- namely, the Kac-Moody currents above.
Put another way, we propose that the Kac-Moody
currents have two interpretations:  one interpretation as the Noether
current of $G$ (with a central extension in the algebra), and another
interpretation as the Noether current of $\tilde{G}_1$ (in which we
propose that the central
term be interpreted in terms of the $B$ field transformations in $\tilde{G}_1$,
reflecting the phases picked up by states in canonical quantization).

As evidence for the interpretation above, we observe that
a Noether current for the $B$ field transformations
should be a two-form current, which would dualize in two dimensions to
a scalar.  We propose therefore that the central extension term in the Kac-Moody
algebra is the Hodge dual of that two-form current, and that the
entire Kac-Moody algebra
be understood as the Noether current for $\tilde{G}_1$ multiplication. 
(The fact that the central extension is dually described in terms of
an anomaly, will be interpreted later in this paper as an example
of anomalies transmuting a classical group symmetry into a higher
group symmetry of the quantum theory.)

To close an intellectual loop, it is natural to conjecture that, for
suitable definitions, representations of
the 2-group $\tilde{G}_1$ should coincide with representations of the
corresponding Kac-Moody algebra, and indeed this conjecture has
been made by others (see {\it e.g.} \cite{urs1}[appendix A], \cite{nganter}).
After all, the states of a WZW model form representations of a Kac-Moody
algebra, and we have also argued that the 2-group $\tilde{G}$ defines a global
symmetry of the theory.
Unfortunately, we are under the impression that the representation theory
of 2-groups is not sufficiently well-developed to address this issue.
(For current work on representations of 2-groups, see for example
\cite{ganterkap,bbfw,ganter-tori}.)

In this language, it is tempting to speculate that level-rank duality should
be realized as some sort of equivalence, perhaps
a Morita equivalence \cite{northana}, of 2-groups, for example as
\begin{displaymath}
\widetilde{SU(n)}_k \: = \: \widetilde{SU(k)}_n ,
\end{displaymath}
where subscripts indicate levels.

Let us conclude this section by observing certain formal similarities
between the structure in this section and our discussion of
Dijkgraaf-Witten boundaries in section~\ref{sect:boundary-dw}.
There, given a bulk Dijkgraaf-Witten theory in three dimensions,
corresponding to a finite $G$ gauge theory, we argued that any boundary
theory should have a $\tilde{G}$ symmetry, where
\begin{displaymath}
1 \: \longrightarrow \: {\bf B} U(1) \: 
\longrightarrow \: \tilde{G} \:
\longrightarrow \: G \: \longrightarrow \: 1
\end{displaymath}
is the extension determined by the same element of $H^3(G,U(1))$
that determined the bulk theory.
In this section, we have studied WZW models.  However, WZW models can be
understood as boundaries of three-dimensional Chern-Simons theories
(see {\it e.g.} \cite{ms-zoo}).  The Chern-Simons theory is a $G$ gauge
theory, classified by an element of $H^3(G,{\mathbb Z})$ (the level),
and we have argued that the boundary WZW theory admits a
$\tilde{G}$ symmetry, where
\begin{displaymath}
1 \: \longrightarrow \: {\bf B} U(1) \: 
\longrightarrow \: \tilde{G} \:
\longrightarrow \: G \: \longrightarrow \: 1
\end{displaymath}
is the extension determined by the same element of $H^3(G,{\mathbb Z})$
that determined the bulk theory.

In the case of discrete torsion in two dimensions and Dijkgraaf-Witten
theory in three dimensions, it was essential that the boundary theory
couple to the bulk theory in such a way that the boundary form potential
transform when the bulk gauge field undergoes a gauge
transformation.  Thus, the parallel would be especially meaningful if
WZW models and Chern-Simons theories coupled in an analogous fashion.
The reference \cite{ms-zoo} restricts to bulk gauge transformations that
are trivial along the boundary, but if we consider more general cases,
it is straightforward to see that
under a gauge transformation, the Chern-Simons action picks up an integral
of a total derivative.  In principle one could imagine that this could
be cancelled by a symmetry transformation of the WZW $B$ field.

In passing, we should mention that such bulk/boundary relationships
have been studied more generally in {\it e.g.}
\cite{lurie-tft,freed-higher,fv} in extended TQFT's, and also
in related work \cite{freed1,freed2,freed3}.  Roughly speaking,
the form of the results is that a ${\bf B}^k G$ gauge theory in
$d$ dimensions has on its boundary a theory with symmetry $\tilde{G}$
given as the extension
\begin{displaymath}
1 \: \longrightarrow \: {\bf B}^{d-2} U(1) \: \longrightarrow \:
\tilde{G} \: \longrightarrow \: {\bf B}^{k} G \: \longrightarrow \: 1 .
\end{displaymath}
As such examples have been discussed extensively elsewhere, and will
play no further role in this paper, we will not discuss them further here.

\subsection{Current algebras}
\label{sect:ex:current-algebras}

We have discussed how Kac-Moody currents in two-dimensional WZW models
can be understood as Noether currents for a symmetry 2-group that
mixes an ordinary group with ${\bf B} U(1)$.  

In this section we shall conjecture an analogous role for current
algebras in four dimensions.

In general, for two currents $J_{\mu,\alpha}$, $J_{\mu,\beta}$
($\mu$ a spacetime Lorentz index), 
their equal-time commutation relations
will have the form \cite{w2}[section 22.6], \cite{iz}[section 11-3]
\begin{displaymath}
[J_{0,\alpha}(\vec{x},t), J_{i,\beta}(\vec{y},t)] \: = \: 
C_{\alpha \beta \gamma} J_{i,\gamma}(\vec{x},t) 
\delta^{d-1}(\vec{x}-\vec{y}) \: + \:
S_{\alpha \beta, i, j} \partial_j \delta^{d-1}(\vec{x}-\vec{y}) ,
\end{displaymath}
where the second term on the right-hand-side is a c-number term known
as the Schwinger term, and we have formally written the expression
for general dimensions $d$, instead of specializing to $d=4$.

The Schwinger terms above have long been interpreted as defining
a projectivization of the algebra of maps from the spacetime into
a classical Lie algebra ${\mathfrak g}$ \cite{faddeev1,fs,mick,dgm}, and 
we propose a refinement of this idea.  Specifically, we propose
that the algebraic structure above be understood as a higher group
extension of the form
\begin{displaymath}
1 \: \longrightarrow \: {\bf B}^{d-1} U(1) \: \longrightarrow \:
\tilde{G} \: \longrightarrow \: G \: \longrightarrow \: 1
\end{displaymath}
generalizing the structure of Kac-Moody algebras.  (For example,
if the four-dimensional spacetime is ${\mathbb R} \times T^3$,
then a projectivization of maps into ${\mathfrak g}$ is the
Lie algebra of a bundle on the triple loop group $L^3 G$, and the triple
loop space of the higher group above is exactly a bundle on $L^3 G$.)
The current for the
${\bf B}^{d-1} U(1)$ piece is a $d$-form, which dualizes in $d$ dimensions
to a scalar.  We conjecture that scalar should be interpreted as
the Schwinger term.  As a consistency check,
note that extensions of the form above are classified by 
homotopy classes of maps\footnote{
For readers unfamiliar with the notation, $[X,Y]$ for $X$, $Y$ spaces or
stacks denotes the homotopy classes of maps from $X$ to $Y$.
}
\begin{displaymath}
[G, B B^{d-1} U(1) ] \: = \: [G, B^d U(1) ] \: = \: 
[G, K({\mathbb Z},d+1)] \: = \: H^{d+1}(G,{\mathbb Z})
\end{displaymath}
as discussed in appendix~\ref{app:class}.
Indeed, in higher-dimensional analogues of WZW models as discussed
in\footnote{
See also \cite{freedglobal} for an updated description relevant for
more general 4-manifolds.  The analysis there interprets these
structures in terms of differential cohomology theories.  A 
thorough description of current algebras along the lines we suggest
should take that into account, but as we are only outlining a conjecture,
we shall
not try to relate those differential cohomologies 
to higher groups in this brief section.  Instead, we leave such an analysis
for future work.
} {\it e.g.} \cite{edglobal}, the analogue of the Wess-Zumino
term in $d$ dimensions is defined by a degree $(d+1)$ form, integrated
over a bounding $(d+1)$-dimensional space, which is precisely consistent
with the classification above.

\subsection{Other Kac-Moody actions}

We have seen that Kac-Moody algebras are closely related to Lie algebras
of 2-groups, and so it is worth pointing out that there are other
actions of Kac-Moody algebras in field theories in dimensions greater than
two.  We list a few examples below.

The work
\cite{cgw,cw,cgsw,dolan-affine} 
described Kac-Moody actions on instanton moduli spaces (see also
\cite{crane1}), albeit at level $0$.  
More recently, other Kac-Moody actions on four-dimensional gauge
theories have been discussed in {\it e.g.} \cite{hms,ac},
though again at level $0$.
As discussed in \cite{vw}[section 4], 
actions of Kac-Moody algebras on cohomology
of moduli spaces of instantons on ALE spaces
were discussed in {\it e.g.} \cite{nak1}.

Examples of hyperbolic Kac-Moody algebra actions on supergravity theories
are discussed in a number of references including
{\it e.g.} \cite{hl1,jul1,jul2,nic1,jul3}.

\section{Cosmological defects and generalized moduli spaces}
\label{sect:cosmo-defects}

\subsection{Generalities on defects and ordinary moduli spaces}
\label{sect:defects:ordinary}

In four dimensional theories, there are a variety of cosmological defects
that are typically classified by the homotopy of the moduli space of
scalar vacuum expectation values.  For example, if ${\cal M}$ denotes
that moduli space, then \cite{vs}[section 3.2]:
\begin{itemize}
\item $\pi_0({\cal M})$ counts components of ${\cal M}$, hence counts
possible domain walls,
\item $\pi_1({\cal M})$ counts possible cosmic strings, determined by
winding in the scalar vevs,
\item $\pi_2({\cal M})$ counts analogues of monopoles, here determined
by noncontractible two-spheres in ${\cal M}$, and
\item $\pi_3({\cal M})$ counts textures.
\end{itemize}

Examples often arise in gauge theories.  Given a $G$ gauge theory with
a Higgs field that breaks $G$ to $H \subset G$, and a suitable nontrivial
potential\footnote{
We are interested in topological defects, not semilocal strings.
In the latter, the potential could vanish identically, so the full moduli
space of Higgs vevs also includes {\it e.g.} vanishing Higgs vevs, or
other vevs for which the unbroken subgroup is different from that above.
}, the moduli space ${\cal M}$ of Higgs vevs can be taken to be
the coset $G/H$.  The homotopy groups of ${\cal M} = G/H$ can be
related to the homotopy groups of $G$ and $H$ using the homotopy long
exact sequence, for example
\begin{displaymath}
\pi_k(G) \: \longrightarrow \: \pi_k(G/H) \: \longrightarrow \:
\pi_{k-1}(H) \: \longrightarrow \: \pi_{k-1}(G) .
\end{displaymath}
We shall see applications of these ideas momentarily.

Now, let us outline
a classification of
analogues of topological defects pertinent to one-form symmetries and 
corresponding two-groups.  
We claim that the relevant analogue of the moduli space\footnote{
Or stack.  See \cite{hs} for a discussion of cases in which the moduli
`space' admits a stack or gerbe structure, and in particular
the relevance to cosmic strings
and so forth of homotopy groups of the resulting gerbe.  We should add
that, although that reference did not make a clear statement regarding
existence of cosmic strings and so forth associated with trivially-acting
finite groups, we have since come to believe that they do exist, 
hence we now believe that homotopy
groups of Deligne-Mumford stacks should indeed have physical relevance.
}
${\cal M}$ of scalar field vevs is the loop space $L {\cal M}$
parametrizing Wilson lines.
After all, Wilson line vevs are sections of a bundle over $L {\cal M}$,
a bundle determined by the ${\bf B} G$ action, in the same way that
for ordinary symmetries, the scalars are sections of a bundle over
${\cal M}$.  
More explicitly, (low-energy) Wilson lines depend upon scalars 
explicitly as, for example,
\begin{displaymath}
P \exp\left( \int \phi^a A_{\mu}^a ds^{\mu} \right),
\end{displaymath}
hence a given Wilson line is determined by a loop in ${\cal M}$, in line with
observations above about cosmic strings, and becomes a (not necessarily
neutral) scalar over the (based) loop space $L {\cal M}$.

Then, just as homotopy of the space of scalar vevs ${\cal M}$
can encode information about topological defects pertinent to an
ordinary group $G$, homotopy of the loop space $L {\cal M}$ of
scalar field vevs can encode information about topological defects
pertinent to ${\bf B}G$.  
To that end, there is a key identity relating the homotopy
of ${\cal M}$ and its based loop space $L{\cal M}$:
\begin{displaymath}
\pi_k({\cal M}) \: \cong \: \pi_{k-1}(L {\cal M}) .
\end{displaymath}

Alternatively, the reader might prefer to think about analogues of
Wilson lines for $BG$ symmetries as defined by Wilson surfaces of
the form
\begin{displaymath}
\exp\left( \int_{\Sigma} F \right)
\end{displaymath}
for $\Sigma$ some two-dimensional submanifold and $F$ a two-form.  Broadly
speaking, these would be associated with elements of $\pi_2({\cal M})$,
but using the identity above,
\begin{displaymath}
\pi_2({\cal M}) \: \cong \: \pi_1(L {\cal M})
\end{displaymath}
and so again we can reduce these considerations to properties of the
loop space.

Let us walk through the implications of this statement for Wilson lines for
one-form symmetries:
\begin{itemize}
\item Analogues of domain walls for Wilson lines should in principle
be counted by $\pi_0(L {\cal M}) = \pi_1({\cal M})$, which correspond
to cosmic strings.  Here the intuition is simply that, as the cosmic
string corresponds to a 't Hooft loop dual to the Wilson line \cite{pk},
then the Wilson line changes as it winds around the cosmic string.
Thus, in this sense, cosmic strings can define analogues of domain
walls for Wilson lines.

For a simple example, consider the case of a $G$ gauge theory broken to
$H \subset G$ by a Higgs vev, as outlined above.  The cosmic strings
are classified by $\pi_1(G/H)$.  Now, if we run a Wilson line
\begin{displaymath}
P \exp\left( \int_C A \cdot ds \right)
\end{displaymath}
around a cosmic string, by taking the curve $C$ to enclose the cosmic
string, then we expect the phase of the Wilson line to be determined
in part by the cosmic string.  In other words, the gauge field along
the Wilson line should couple to a bundle on $C$ induced by the cosmic
string (as a 't Hooft line).  Now, it is not a coincidence that,
at least for $G$ connected and simply-connected, the classification
of cosmic strings in this theory matches the classification of induced
bundles on circles.  Specifically,
from the homotopy long exact
sequence, if $G$ is connected and simply-connected,
$\pi_1(G/H) = \pi_0(H)$.  Now,
principal $H$ bundles on $S^1$ are classified by
\begin{displaymath}
[S^1, BH] \: = \: \pi_1(BH) \: = \: \pi_0(H),
\end{displaymath}
so we see that the classification of cosmic strings matches that of
principal $H$ bundles on a circle, as one would expect from the fact
that the cosmic string (as a 't Hooft loop) is inducing a bundle on
the enclosing circle.  In particular, a cosmic string acts as a domain
wall in the space of Wilson line vevs:  depending upon whether the
curve $C$ encloses a cosmic string (and the number of times it wraps),
one should get a different component of the space of Wilson line vevs.

\item Analogues of cosmic strings for Wilson lines should in principle
be counted by $\pi_1(L {\cal M}) = \pi_2({\cal M})$, which correspond
to monopoles.  To help explain why this is sensible, first recall that
a cosmic string can be interpreted as a 't Hooft line \cite{pk}.
A charged particle that walks along a loop enclosing
the 't Hooft line will pass through a transition function for a nontrivial
bundle generated by the 't Hooft line.  The analogue of a 't Hooft line
for a $BG$ symmetry is a monopole, which generates a nontrivial
$G$-gerbe.  Indeed, as one drops a monopole through a Wilson line,
its vev will change\footnote{
See for example \cite{bry-bk}[section 7.1], but in fact at some level this is
merely an unwinding of the definition of a monopole across the surface of an
enclosing $S^2$.
}, so in this fashion one has
a notion of winding for Wilson lines. 
A Wilson line walking through a loop in which it wraps a monopole is
schematically illustrated in figure~\ref{fig:1loop:monopole}.

\begin{figure}[h]
\centering
\begin{picture}(100,110)(0,0)
\Oval(95,50)(20,25)(0)
\Oval(60,25)(20,10)(0)
\Oval(60,75)(20,10)(0)
\Oval(25,50)(20,25)(0)
\DashArrowArc(60,50)(50,40,60){3}
\DashArrowArc(60,50)(50,300,320){3}
\DashArrowArc(60,50)(50,120,140){3}
\DashArrowArc(60,50)(50,220,240){3}
\Vertex(95,50){5}
\end{picture}
\caption{A Wilson line about a closed loop itself moves through a closed
path, which encloses a monopole (right side).
}
\label{fig:1loop:monopole}
\end{figure}
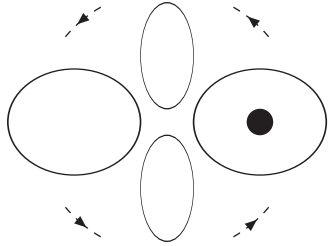

As a particular example, let us consider analogues of Alice strings.
Recall (see {\it e.g.} \cite{vs}[section 4.2.4], \cite{s1,s2,pk,blp}),
Alice strings arise in $SO(3)$ (and $SU(2)$) gauge
theories with matter in the 5-dimensional representation,
consisting of real symmetric traceless $3 \times 3$ matrices.  A typical
Higgs vev, {\it e.g.}
\begin{displaymath}
\left[ \begin{array}{ccc}
1 & 0 & 0 \\
0 & 1 & 0 \\
0 & 0 & -2 \end{array} \right]
\end{displaymath}
has stabilizer
$S(O(2)\times O(1)) = O(2) \subset SO(3)$, so with a suitable
potential
forcing the vacua to lie on the coset,
the moduli space of Higgs vevs is
$SO(3)/O(2) \: = \: {\mathbb R}{\mathbb P}^2$.
(Equivalently, we could work in an $SU(2)$ gauge theory, in which
case the stabilizer of the Higgs vev above would be\footnote{
Pin$(2)$ here can be understood as the
normalizer of a maximal torus.  Pin$(2)$ is also a double-cover of
$O(2)$.
}
${\rm Pin}(2) \subset SU(2)$, and again
$SU(2)/{\rm Pin}(2) = {\mathbb R}{\mathbb P}^2$.)
This moduli space has $\pi_1 = {\mathbb Z}_2$, hence there are
cosmic strings in this model, in which the Higgs field vev wraps
nontrivially around the moduli space.  More to the point, the cosmic
strings are classified by ${\mathbb Z}_2$.  The Wilson lines act
on electric charges of the unbroken $O(2)$ (Pin$(2)$ in the $SU(2)$
theory) by flipping their signs:  representations of $O(2)$ (resp. Pin$(2)$)
are pairs of matched $SO(2)$ (resp. Spin$(2)$) representations of
opposite sign \cite{tung}[section 11.1], and the Wilson line about the string,
which takes values in the diconnected
component, flips the signs, exchanging the two $SO(2)$ (resp. Spin$(2)$)
representations.  The fact that electric charges undergo a sign flip
under parallel transport about the string, ultimately a consequence
of the fact that the strings are classified by ${\mathbb Z}_2$,
is the defining characteristic\footnote{
As an aside for interested readers, let us discuss another characteristic
feature of Alice strings in this language.
Take a $SO(2)$ charge $q$, and pass through the middle of a pair of parallel
Alice strings.  Asymptotically, the pair of parallel Alice strings yield
an $O(2)$ bundle whose disconnected component is trivializable, so $SO(2)$
charges are sensible asymptotically (though not invariant under $O(2)$ gauge
transformations).  Pass a charge $q$ through the middle of the pair, and
it will come out on the other side as charge $-q$, hence to preserve charges,
the pair of Alice strings are then interpreted as having charge $+2q$.
This phenomenon is known as `Cheshire charge.'
The basic issue in all these cases seems to be one of ill-fated attempts to
define $SO(2)$ matter in an $O(2)$ gauge theory.
} of Alice strings.

The analogue of Alice strings for Wilson lines would involve
a moduli space ${\cal M}$ such that $\pi_1( L {\cal M}) = \pi_2({\cal M})
= {\mathbb Z}_2$.  In general, for a $G$ gauge theory with a potential
forcing all the Higgs vevs to break $G$ to the same subgroup $H \subset G$,
the moduli space of Higgs vevs is the coset $G/H$.  For example\footnote{
We would like to thank A.~Knutson for suggesting this example.
},
consider an $SU(3)$
gauge theory in which a Higgs vev breaks the $SU(3)$ to $SO(3) \subset SU(3)$.
This would happen for a Higgs field in the ${\rm Sym}^2 {\bf 3} = 
{\bf 6}$ representation,
with a vev given by the identity.
This theory
has the property that
\begin{displaymath}
\pi_2( SU(3) / SO(3) ) \: = \: {\mathbb Z}_2 .
\end{displaymath}
In particular, a Wilson line following a path that swoops around a monopole,
as in figure~\ref{fig:1loop:monopole}, would undergo a gauge transformation
in the nontrivial element of ${\mathbb Z}_2$.

\item Analogues of monopoles for Wilson lines should in principle
be counted by $\pi_2(L {\cal M}) = \pi_3({\cal M})$, which correspond
to textures.  
\item Finally, analogues of textures for Wilson lines should in principle
be counted by $\pi_3(L {\cal M}) = \pi_4({\cal M})$.
\end{itemize}

Another standard matter involves defects ending on other defects.
For example, suppose a gauge symmetry $G$ is broken in two steps at
successively lower energies, first to a subgroup $H \subset G$, and
then $H$ is completely broken.  (See {\it e.g.} \cite{urs-thesis}[section 5.7.1]
for a longer review.)
\begin{itemize}
\item  {\bf Domain walls ending on strings:}
After the first step, when $G$ is
broken to $H$, the theory will
have cosmic strings counted by $\pi_1(G/H)$.  After the second step,
the theory will have domain walls counted by $\pi_0(H)$.  The domain walls
of the second step can end on the strings of the first step, consistent
with the observation that (for $G$ connected and simply-connected),
$\pi_1(G/H) \cong \pi_0(H)$.
\item {\bf Strings ending on monopoles:}
After the first step, when $G$ is broken to $H$, the theory will have
monopoles counted by $\pi_2(G/H)$.  After the second step, the theory
will have strings counted by $\pi_1(H)$.  The strings of the second step
can end on the monopoles of the first step, consistent with the observation
that (for $G$ 1- and 2-connected), $\pi_2(G/H) \cong \pi_1(H)$.
\end{itemize}

The analogue for Wilson lines is similar:
\begin{itemize}
\item {\bf Domain walls ending on strings:}
After the first step, when $G$ is broken to $H$, the theory will have cosmic
strings counted by $\pi_1(L G/H) = \pi_2(G/H)$.  After the second step,
the theory will have domain walls counted by $\pi_0(L H) = \pi_1(H)$.
Thus, the analogue of domain walls ending on strings, can equivalently
be understood as ordinary strings ending on ordinary monopoles.
\end{itemize}

In principle, the same analysis can be continued for higher
$p$-form symmetries.  For example, analogues of cosmological defects
for a Wilson surface corresponding to a 2-form symmetry would be
measured by $\pi_k(L^2 {\cal M}) = \pi_{k+2}({\cal M})$.
For example, a domain wall for a Wilson surface would correspond to an ordinary
monopole:  the set of values for a Wilson surface would break into
components, determined by how many times the Wilson surface wraps the
monopole.  

In a similar fashion, analogues of cosmological defects
for a Wilson surface corresponding
to a $p$-form symmetry would be measured by 
$\pi_k(L^p {\cal M}) = \pi_{k+p}({\cal M})$.
Of course, for a spacetime of a given dimension, there is an upper limit
to the degree of homotopy groups that can be realized in that spacetime.
The intuition behind higher examples rapidly becomes obscure,
so we will not pursue this further, except to note that this does provide
a physical interpretation to higher homotopy groups of the moduli space
${\cal M}$ of scalar field vevs.

\subsection{Generalized moduli spaces}
\label{sect:genl-moduli-spaces}

So far we have discussed cosmological defects and higher loops for
ordinary moduli spaces.  Next, we shall consider cases in which a higher
group action exists on the moduli space of the field or string theory.
In examples in which the moduli `space' admits an action
of a higher group, the `space' is a generalized space known as
a stack.  

The paper \cite{hs} studied a number of properties of four-dimensional
theories whose moduli `spaces' were of this form.  For completeness,
we begin by reviewing the highlights of cosmological defects studied in
\cite{hs}, here.  We will then turn to other moduli `spaces' appearing
in field and string theory, motivated by our discussion of higher group
actions on WZW models and current algebras.

To begin,
consider as a prototypical example a four-dimensional $U(1)$ gauge theory with
$2$ chiral multiplets of charge $+k$ and $2$ chiral multiplets of charge
$-k$, for $k > 1$, along with massive minimally-charged matter.
As discussed earlier in section~\ref{sect:gerbes} as well as {\it e.g.}
\cite{nr,msx,glsm,hhpsa,hs}, 
both the quantum field theory and the moduli `space' 
admit an action of ${\bf B} {\mathbb Z}_k$,
and so its spectrum of cosmological defects is slightly different than
for ordinary cases.

Let ${\cal M}$ denote the moduli `space' of the theory above,
and $M$ the ordinary moduli space of the corresponding theory with $k=1$.
As discussed in \cite{hs}, there is a long exact sequence of 
homotopy groups
\begin{displaymath}
\cdots \: \longrightarrow \: \pi_m({\bf B} {\mathbb Z}_k) \: \longrightarrow \:
\pi_m({\cal M}) \: \longrightarrow \: \pi_m(M) \: \longrightarrow \:
\pi_{m-1}({\bf B} {\mathbb Z}_k) \: \longrightarrow \: \cdots .
\end{displaymath}

In principle, one would expect cosmological defects in the theory with
$k > 1$ to be counted by\footnote{
This assumes that {\it e.g.} cosmic strings defined by nontrivial bundles
for trivially-acting groups exist and are different from ordinary cosmic
strings, in other words that $\pi_1({\bf B} {\mathbb Z}_k)$ is counting
cosmic strings of this type.  At the time reference~\cite{hs} was written,
this matter was ambiguous, but now seems to have been settled in the
affirmative.  Such cosmic strings are defined by nontrivial
${\mathbb Z}_k$ bundles over enclosing circles, but in which the
${\mathbb Z}_k$ acts trivially on the other fields of the theory.
(In particular, a map from any space $X$ into ${\bf B} {\mathbb Z}_k$ is
defined by a ${\mathbb Z}_k$ bundle on $X$, both for the classifying space
and for the stack, which is part of the reason why cosmic strings on
such higher groups would have the form described.)
} $\pi_m({\cal M})$, rather than
$\pi_m(M)$.  In the case above, the theory with $k=1$ admits only
monopoles, as only $\pi_2(M) \neq 0$.  The corresponding theory with
$k > 1$ also admits monopoles, but the counting is slightly different.
In particular, since $\pi_2({\bf B} {\mathbb Z}_k)$ vanishes, the homotopy
sequence above reduces to
\begin{displaymath}
0 \: \longrightarrow \: \pi_2({\cal M}) \: \longrightarrow \:
\pi_2(M) (\cong {\mathbb Z}) \: \longrightarrow \:
\pi_1({\bf B} {\mathbb Z}_k) ( \cong {\mathbb Z}_k) \: \longrightarrow \: 0 .
\end{displaymath}
Thus, our theory with $k>1$ also admits countably many monopoles, but
counted slightly differently than in the case $k=1$.  In particular,
we see that not every monopole in the $k=1$ theory lifts to a monopole
in the $k>1$ theory -- apparently only monopoles whose charges satisfy
a divisibility constraint lift to the $k>1$ theory, as one might expect
on general principles of charge quantization.

As one more example, consider a four-dimensional theory in which
one gauges a finite group $G$, all of which acts trivially on the theory.
If the ungauged theory had moduli space $M$, then the gauge theory has
moduli `space' $M \times {\bf B} G$.
In this case, the counting is again modified, but in a more trivial
fashion, as
\begin{displaymath}
\pi_1(M \times {\bf B}G) \: = \: \pi_1(M) \oplus \pi_1({\bf B}G)
\end{displaymath}
and other homotopy groups are unmodified.
Again the counting of cosmological defects is modified, but in a much
more trivial fashion.

So far we have considered moduli `spaces' of four-dimensional field and
string theories admitting an action of ${\bf B}G$ for $G$ finite.
Mathematically, there is a notion of generalized `spaces' that would admit
actions of ${\bf B}G$ for $G$ nonfinite; however, it is not known at
present how to, for example, define a sigma model on such.  (Technically,
it is known how to define a sigma model on a Deligne-Mumford stack,
but not an Artin stack.)  In the remainder
of this section, we will explore the possibility of whether such more
general moduli `spaces' might appear in physics via a simple example.
We will get rather confusing results, suggesting that possibly the only
generalized moduli `spaces' relevant for physics are of the former 
(Deligne-Mumford) type.

In section~\ref{sect:ex:wzw} we reviewed existing results on how WZW models
are believed to admit an action of $\tilde{G}$, an extension of a 
Lie group $G$ (partially defining the WZW model) by ${\bf B} U(1)$, where the
extension class is determined by the level of the WZW model.  We could imagine
fibering the WZW model over some other space $X$, as described in
{\it e.g.} \cite{ds,gates1,gates2,gates3,gates4,gates5}.  The resulting
two-dimensional theory has a semiclassical moduli `space' given by a 
$\tilde{G}$-bundle over $X$, another example of a stacky moduli space,
albeit an Artin stack and not a Deligne-Mumford stack.

The two-dimensional example above would, for obvious reasons, not admit
a wide variety of cosmological defects, but formally we can imagine
higher-dimensional analogues.  Following the proposal in 
section~\ref{sect:ex:current-algebras} for an interpretation of
current algebras in terms of higher groups, one could similarly imagine
fibering a higher-dimensional current algebra over some space $X$,
at least in low-energy effective field theory.  If $\tilde{G}$ is the
higher symmetry group of the current algebra, then by analogy with 
fibered two-dimensional cases, one would have a moduli `space' given by
a $\tilde{G}$-bundle over $X$.

In such a case, since $\tilde{G}$ was determined by an extension encoding
an anomaly, in effect some of the homotopy groups of $\tilde{G}$ and hence
of the moduli stack would be determined by an anomaly.  Put another way,
one would have anomaly-induced cosmological defects.

For example, let us consider the $SU(3)$ current algebra described in
\cite{edglobal}.  In present language, section~\ref{sect:ex:current-algebras} 
conjectures that it should
have a $\tilde{G}$ symmetry, where $\tilde{G}$ is given by
\begin{displaymath}
1 \: \longrightarrow \: {\bf B}^3U(1) \: \longrightarrow \: \tilde{G} \:
\longrightarrow \: SU(3) \: \longrightarrow \: 1 ,
\end{displaymath}
where the level $k$
of the current algebra ($n$ in the notation of \cite{edglobal})
determines the extension class.  We can consider this model by itself,
in effect fibering over a point. 

From the long exact homotopy sequence (which operates for higher
groups in essentially the same way as for spaces \cite{urspriv}),
we see that for $m \neq 4, 5$,
$\pi_m(\tilde{G}) = \pi_m(SU(3))$, which (for $n \neq 4, 5$) is nonzero
for $n=3$ and $n=7$ for example.  For $m = 4, 5$, there is a long exact sequence
\begin{displaymath}
0 \: \longrightarrow \: \pi_5(\tilde{G}) \: \longrightarrow \:
\pi_5(SU(3)) (\cong {\mathbb Z}) \: \longrightarrow \:
\pi_4({\bf B}^3 U(1)) (\cong {\mathbb Z}) \: \longrightarrow \:
\pi_4(\tilde{G}) \: \longrightarrow \: 0 .
\end{displaymath} 
The map $\pi_5(SU(3)) \rightarrow \pi_4({\bf B}^3U(1))$ should be
proportional to the level $k$ of the current algebra, so that at level zero,
the sequence splits, and $\pi_5(\tilde{G}) \cong \pi_5(SU(3))$.
If the level $k \neq 0$, then $\pi_5(\tilde{G}) = 0$ and
$\pi_4(\tilde{G}) = {\mathbb Z}_k$.

Unfortunately, it is not clear how to interpret the
homotopy groups of the Artin stack above physically.  It is possible
that one must perform a more subtle analysis along the lines of
\cite{freedglobal}.  However, for the moment we will instead
interpret this to mean that the physical relevance of Artin stacks
is unclear, and when computing {\it e.g.}
cosmological defects, one should restrict to cases involving
Deligne-Mumford stacks (meaning, actions of ${\bf B}G$ for $G$
finite).

\subsection{Analogues of Goldstone's theorem and generalized moduli spaces}

Previously in section~\ref{sect:review:nati} and in \cite{gksw}, 
analogues of Goldstone's
theorem were discussed for counting Goldstone bosons in theories
with spontaneously broken higher group symmetries.
For example, the Goldstone boson associated with a spontaneously broken
${\bf B}_{\flat} U(1)_{\rm conn}$ in an abelian gauge theory was
proposed in \cite{gksw} to be the gauge field itself.

Now, for ordinary groups, there is additional content in Goldstone's
theorem.  In principle, Goldstone's theorem not only gives a count of
bosons associated with a spontaneously broken global symmetry, but also
describes the local dimension of the moduli space.  After all, 
if a global symmetry $G$
is broken to a subgroup $H$, then one has dim $G/H$ massless bosons
ultimately because the moduli space of vacua locally looks like $G/H$.

We have already discussed how generalized moduli `spaces' (stacks) can
admit actions of higher groups, so it is natural to ask what analogous
statements can be made for the moduli stacks appearing in such gauge
theories.  For example, if one spontaneously breaks a ${\bf B}G$ symmetry
to a subgroup
${\bf B}H$, then does a coset of the form
${\bf B}G/{\bf B}H$ play a role in physics, in the same way that for
ordinary groups, cosets $G/H$ play a role in understanding moduli spaces
in spontaneous symmetry breaking?
In this section, we will not reach any firm conclusions,
but we will discuss these issues in simple examples.

Let us first outline a little pertinent 
mathematics.
If $\tilde{G}$ is a higher group,
and $\tilde{H}$ a sub(higher)group,
then the homotopy quotient $\tilde{G}/\tilde{H}$ exists \cite{urspriv}, 
and there is a long exact
sequence of homotopy groups 
of exactly the same form as for ordinary groups \cite{urspriv}, namely
\begin{displaymath}
\pi_k(\tilde{G}) \: \longrightarrow \: \pi_k(\tilde{G}/\tilde{H}) 
\: \longrightarrow \:
\pi_{k-1}(\tilde{H}) \: \longrightarrow \: \pi_{k-1}(\tilde{G}) .
\end{displaymath}
For the higher groups ${\bf B}^k G$, we can use the fact that
\begin{displaymath}
\pi_m( {\bf B}^k G) \: = \: \pi_{\pi_{m-1}}( {\bf B}^{k-1} G) \: = \:
\cdots \: = \:
\pi_{m-k}(G)
\end{displaymath}
to relate the homotopy groups of ${\bf B}^k G$ to those of $G$.
The `spaces' of such higher groups can be understood as stacks,
or `generalized spaces' in the language of the previous section.

Second, it may be pertinent to distinguish two cases.  As previously
outlined, stacks associated
with ${\bf B}G$ for $G$ finite are different both mathematically and physically
from those associated with ${\bf B} G$ for $G$ non-finite.
For the former, known technically as Deligne-Mumford stacks, 
it is known how to define
a sigma model \cite{msx}.  For the latter, known technically as Artin stacks, 
the physics
is largely unknown, and we suspect likely to be more subtle than
for Deligne-Mumford stacks.  As a result, it is entirely possible that physical 
interpretations of homotopy groups of gerbe structures may differ
between such cases.

Now, let us consider a concrete example.  As was reviewed in
section~\ref{sect:review:nati}, 
reference~\cite{gksw} discusses abelian gauge theories as
examples of theories with ${\bf B}_{\flat} U(1)_{\rm conn}$ actions.
As discussed there, in some dimensions, this symmetry is spontaneously
broken, and the abelian gauge field is identified with the Goldstone
boson.
Let us try to interpret this result in terms of possible moduli stack
structures.
\begin{itemize}
\item First, let us proceed naively, and look for an interpretation
of the homotopy groups of ${\bf B}_{\flat} U(1)_{\rm conn}$ itself.
Now, as outlined in section~\ref{sect:overview},
the homotopy groups of ${\bf B}_{\flat} U(1)_{\rm conn}$ are the
same as those of $K(U(1),0)$, which is to say,
\begin{displaymath}
\pi_0( {\bf B}_{\flat} U(1)_{\rm conn} ) \: = \: U(1), \: \: \:
\pi_n( {\bf B}_{\flat} U(1)_{\rm conn} ) \: = \: 0 \mbox{ for }
n > 0 .
\end{displaymath}
However, it is not entirely clear to us how these homotopy groups
would be interpreted in an abelian gauge theory.
\item Next, let us take a slightly more radical approach.
Since we have an abelian gauge theory, it is tempting to replace the
moduli space of the theory $M$ with an Artin stack, a $U(1)$ gerbe
over $M$ with fibers ${\bf B}U(1)$.  This would be in the spirit, though
not the detailed method, of \cite{glsm,msx,hs}:  a sigma model on
a stack of this form should morally
involve a path integral over $U(1)$ bundles with
connection, hence an abelian gauge theory, though previous work has only
made claims about stacks involving finite gerbes (Deligne-Mumford stacks).
Sigma models on Artin stacks have not been studied at all, and are likely
to be subtle to interpret.  That said, if this extremely naive approach
were to hold water, then in principle we could effectively encode
the $U(1)$ gauge fields in the `stacky' structure of the moduli space,
consistent with the naive application of a generalized Goldstone's
theorem as giving
${\bf B}U(1)$ directions in the moduli `space'
in the spontaneously broken symmetry phase.  That said, very naively,
similar statements could also be made in other dimensions, even in which
the large-loop behavior changes and one no longer has an IR interpretation
in terms of spontaneous symmetry breaking.
\end{itemize}

At this point in time, we are therefore not able to make any
firm statements.

\section{Anomalies as transmutation}
\label{sect:anom:transmutation}

\subsection{Two-dimensional cases}

Let us now turn to the application of 2-groups to anomalies in 
two-dimensional quantum field theories.
If $J$ is a Noether current, or rather the corresponding composite operator
in some two-dimensional quantum field theory,
then the $J^2$ anomaly in two dimensions is
well-known to be proportional to  ${\rm Tr}\, F \wedge F$, involving
the curvature of a coupled (possibly nondynamical) gauge field.
Anomalies of this sort
are indicative of curvature on the space of gauge transformations,
or somewhat more precisely, indicate
that the path integral measure
fails to be well-defined under families of gauge transformations.
(See for example \cite{agg}[section 3] for a pertinent discussion.)
This can happen even if there are no chiral fermi
zero modes:  the path integral measure can be invariant under a single
gauge transformation, but not well-behaved in families.

The fact that the anomaly arises in families of gauge transformations
suggests an interpretation of the
anomaly.  Although, the anomaly `breaks' the classical symmetry
group $G$, 
the form in which it is
broken suggests that the quantum theory might have a different symmetry
`group,' perhaps one in which associativity is broken up to higher cocycles,
where the associators are encoded implicitly in the phases one picks up
around loops.

In this section, we will propose that in such cases,
the quantum theory has a symmetry 2-group, instead of
a symmetry group, specifically
the 2-group
$\tilde{G}_1$ defined in \cite{sp} by an extension
\begin{displaymath}
1 \: \longrightarrow \: {\bf B} U(1) \: 
\longrightarrow \: \tilde{G} \:
\longrightarrow \: G \: \longrightarrow \: 1 .
\end{displaymath}
In particular, in this interpretation,
the anomaly breaks the classical symmetry, but
simultaneously encodes an extension of $G$ to a 2-group symmetry,
and so describes a transmutation of a zero-form symmetry into
a 2-group symmetry, a structure we have seen elsewhere in this paper.

One way to see this structure is via a dual representation of
the fermions, in a WZW model.
As is well-known \cite{ed-wzw}, the nonabelian bosonization of fermions with
classical symmetry group $G$ is described by a WZW model,
and we described previously in section~\ref{sect:ex:wzw}
how WZW models form examples
of theories in which ordinary group actions are replaced
with 2-group actions.

In this language, the fermion chiral symmetry currents correspond to,
in the WZW model,
classically the left and right multiplications by $G$.  These classical
symmetries are modified in the quantum theory by the central extension
term in the Kac-Moody algebra, 
which, as has been argued here and elsewhere, suggests they should be
extended to left and right multiplication by the 2-group
$\tilde{G}_1$, defined by the extension of $G$ by ${\bf B}U(1)$
dictated
by the level of the WZW model.   
The fact that
the 2-group $\tilde{G}_1$ explicitly appears in the bosonized
representation of the fermions, argues that it must also appear
in the quantized fermion theory, and since in the WZW model the 2-group
$\tilde{G}_1$ extends the classical group $G$, in the dual
fermion theory the 2-group $\tilde{G}_1$ must replace the anomalous
action of the classical symmetry group $G$.

Another check of this proposal arises when we couple the fermions to
a gauge field.
In principle, if the classical symmetry group becomes in the quantum
theory a symmetry 2-group, then to gauge $G$,
any classical principal $G$ bundle $P$
must be lifted to a principal $\tilde{G}_1$ bundle.
The obstruction to this lift\footnote{
Obstructions to spin structures may be more familiar to some readers,
and are analogous \cite{brymc1}[section 3].  Given
a principal $SO(n)$ bundle $P$ over some manifold $M$,
as is well-known there is an obstruction to lifting to a principal
Spin$(n)$ bundle, measured by the second Stiefel-Whitney class $w_2(P)$.
Corresponding to this obstruction is a ${\mathbb Z}_2$ gerbe on $M$,
whose characteristic class is given by $w_2(P) \in H^2(M,{\mathbb Z}_2)$.
} is discussed
in \cite{sp}; it is of the form
\begin{displaymath}
k \, {\rm Tr}\, F \wedge F ,
\end{displaymath}
where $F$ is the curvature of $P$ and
$k \in H^3(G, {\bf Z})$ is the level of the affine algebra,
or equivalently the cohomology class describing $\tilde{G}_1$ as
an extension of $G$.  (For example, for a trivial extension, $k=0$,
and there is no obstruction.)  Demanding that this obstruction vanish
for all bundles $P$ is, in general, not possible, and so gauging
the classical symmetry is not possible, as one expects for
an anomalous symmetry.

Another check arises in fibered WZW models
\cite{ds,gates1,gates2,gates3,gates4,gates5}, describing
{\it e.g.} a bosonization of the fermions in a nonlinear sigma model.
The fibering of the WZW model is described by a principal $G$ bundle $P$
over the target space.
Given a principal $G$ bundle $P$, there is an obstruction to lifting it
to a principal $\tilde{G}_1$ bundle.  For the same reasons
as above, that obstruction is
\begin{displaymath}
k \, {\rm Tr}\, F \wedge F ,
\end{displaymath}
where $k \in H^3(G, {\bf Z})$ is the level of the WZW model,
or equivalently the cohomology class describing $\tilde{G}_1$ as
an extension of $G$.  (For example, for a trivial extension, $k=0$,
and there is no obstruction.)
In a heterotic theory,
this obstruction corresponds to a contribution
to the Green-Schwarz anomaly in a (0,2) theory describing a left-moving
current algebra at level $k$, for example.

Consider more general fibered WZW models, in which one bosonizes both left
and right-movers.  Suppose one has classically a principal
$G_L \times G_R$ bundle, with different groups for left and right-movers.
If we call this bundle $P$, then anomaly cancellation in this case
becomes the statement that the $G_L \times G_R$ bundle can be lifted to a
$\widetilde{ G_L \times G_R }_1$ bundle; obstructions on either side
separately are cancelled out when one takes both factors together.

The idea that a classical group can transmute into a 2-group
symmetry of the quantum theory, as we have
formulated it above, may seem odd, but
analogous transmutations may also sometimes appear along
renormalization group flows.  For example, in two dimensions it is a common
result that global symmetry groups become enhanced at IR fixed points
to affine algebras, which as previously discussed are linked to
2-groups.  In this language, at least naively RG flow would appear
to describe a transmutation from a (UV) global symmetry group
to an (IR) 2-group.

\subsection{Conjectures on elliptic genera}

We can gain further possible insight into the physical relevance of 2-groups by
considering elliptic genera.  The basic Witten genus describes a quantum
theory of a set of fermions with a Spin rotation symmetry, fibered
over some space $X$, coupled via the tangent bundle.

In discussions of elliptic genera, it is well-known that a topological
group named String plays an important role.  For this discussion,
we should point out that
String$(n)$ can be built as a 2-group, extending Spin$(n)$ at level one by
${\bf B}U(1)$
\cite{sp}[section 4.3], \cite{kw}[section 7], \cite{bcss}:
\begin{displaymath}
1 \: \longrightarrow \: {\bf B} U(1) \: \longrightarrow \:
{\rm String}(n) \: \longrightarrow \: {\rm Spin}(n) \:
\longrightarrow \: 1 .
\end{displaymath}

With the above in mind, it is very tempting to conjecture that the
states appearing in elliptic genera form a representation of
the 2-group String$(n)$.  In particular, the states lie in
representations of an affine extension of $so(n)$, and as we discussed
in section~\ref{sect:ex:wzw},
it is tempting to identify representations of the 2-group and the
affine algebra.  However,
as we discussed there, the representation theory
of 2-groups does not seem to be sufficiently well-developed at the
moment to completely answer this question one way or another, so we
leave this matter for future work..

\subsection{Higher-dimensional conjectures}

In $2n$ dimensions, anomalies can arise from higher-dimensional
analogues of the same issue with well-definedness in families.
As discussed in {\it e.g.} \cite{agg}[section 3], consider a
theory of
charged fermions on $S^{2n}$,
coupling to a principal $G$ bundle $P$ over some space $X$.
Under a one-parameter family of gauge transformations,
the fermion path integral may fail to be well-defined, measured
at least in part by an element of $H^{2n+1}(G)$.

Proceeding solely by analogy with the two-dimensional case discussed
above,
in our language, this might correspond to the
statement that the correct symmetry group of the theory is
a $2n$-group, which topologically is a $(2n-1)$-gerbe over $G$,
an extension of $G$ by ${\bf B}^{2n-1} U(1)$.
If we take enough loops,
this becomes a central extension of $L^{2n-1} G$ by $U(1)$.
It is natural to conjecture that obstructions to lifting principal $G$
bundles to principal $\tilde{G}_1$ bundles, for $\tilde{G}_1$ a higher group
extension as above, are characterized by a degree $2n+2$ characteristic
class, of the form
\begin{displaymath}
k \, {\rm Tr}\, F \wedge \cdots \wedge F
\end{displaymath}
($n+1$ factors), which would certainly tie into the description of
anomalies, as well as central extensions in higher dimensional current
algebras as outlined in section~\ref{sect:ex:current-algebras}.
We will leave this matter for future work.

\section{Conclusions}

In this paper we described the $p$-form symmetries of
\cite{gksw} as special cases of higher group actions in quantum field
theory.  After briefly outlining the definition of higher groups,
we outlined several examples, both reviewing some already in the literature
as well as outlining some new examples.  We discussed the role of
`generalized moduli spaces' in quantum field theory, admitting actions of
higher groups, and discussed cosmological defects in this context.
We concluded with a proposal for an interpretation of certain anomalies
in terms of higher group symmetries of quantum theories.

There is also evidence for a role for 2-groups in moonshine,
see {\it e.g.} \cite{nora09}, \cite{ganter-tori}[section 5].
It has also been suggested that such higher groups
might play a further role in
the understanding of Mathieu groups in SCFT's as described
in {\it e.g.} \cite{eot-m,eguchi1,eguchi2,wendland1,wendland2}, as for
example the Mathieu group $M_{12}$ is a subgroup of a
groupoid denoted $M_{13}$ \cite{nganter,ecm}.

In section~\ref{sect:gerbes}, we discussed two-dimensional gauge theories in
which a finite subgroup $G$ of the gauge group acts trivially on the
massless matter.  We argued that at low energies, these theories have
a $BG$ symmetry, and outlined how these theories decompose into disjoint
unions of simpler theories.
It would be interesting to understand if there are analogues of decomposition
for
any notion of $BG$ gauge theories or other higher gauge theories in
any dimension.

It is tempting to speculate that some of the relations we have explored
between ordinary groups and higher groups may have analogues outlined
in \cite{ms-old}, but we shall leave such considerations for future work.

\section{Acknowledgements}

We would like to thank M.~Ando, N.~Aramian,
N.~Epa, D.~Freed, N.~Ganter, A.~Knutson,
T.~Pantev, U.~Schreiber, and R.~Thorngren for useful conversations.
This paper began as an joint effort with M.~Ando, who provided initial
ideas and commentary, and N.~Ganter and N.~Epa similarly contributed to the
development at an intermediate stage.  U.~Schreiber contributed many
references and results on higher groups, and N.~Aramian also contributed
many technical observations.
Although we were motivated to write up our results by the
publication of \cite{gksw}, this paper is based on discussions
of applications of 2-groups to physics that took place over several years,
during which time E.S.
was supported by NSF grants DMS-0705381, PHY-0755614, PHY-1068725,
and PHY-1417410.

\appendix

\section{Topological classification of extensions}
\label{app:class}

In this paper we have encountered a number of higher groups $\tilde{G}$
constructed
as extensions of the form
\begin{displaymath}
1 \: \longrightarrow \: {\bf B}^k U(1) \: \longrightarrow \: 
\tilde{G}_{1, k} \: \longrightarrow \: G \: \longrightarrow \: 1 ,
\end{displaymath}
\begin{displaymath}
1 \: \longrightarrow \: {\bf B}^k U(1)_{\rm conn} \: \longrightarrow \: 
\tilde{G}_{c, k} \: \longrightarrow \: G \: \longrightarrow \: 1 ,
\end{displaymath}
\begin{displaymath}
1 \: \longrightarrow \: B^k U(1) \: \longrightarrow \: 
\tilde{G}_{0, k} \: \longrightarrow \: G \: \longrightarrow \: 1 ,
\end{displaymath}
for some other (higher) group $G$.  To classify the possible extensions,
we first need to compute the topological classes of extensions.

An extension of the form above is the total space of a ${\bf B}^k U(1)$
(or ${\bf B}^k U(1)_{\rm conn}$, or $B^k U(1)$)
bundle over $G$.  Regardless of our version of $B^k U(1)$, for the purposes
of a topological classification of bundles, we can replace it with
a suitable Eilenberg-Maclane space:
\begin{displaymath}
B^k U(1) \: = \: K({\mathbb Z},k+1), \: \: \:
B U(1) \: \cong \: {\mathbb C}{\mathbb P}^{\infty} .
\end{displaymath}
(In particular, since the geometric realizations of all three options are
homotopic to one another\footnote{
For $\tilde{G}_{1,k}$ and $\tilde{G}_{0,k}$, this follows from 
\cite{urs-thesis}[theorem 4.4.36].  For $\tilde{G}_{c,k}$, this follows
from unpublished work of D. Pavlov \cite{urspriv}.
The reader should also note that the geometric realization of
${\bf B}^k U(1)_{\rm conn}$ forgets the differential forms data, and so
loses quite a bit of information.
},
the resulting topological classification will
not depend upon whether we are extending by ${\bf B}^k U(1)$ or
${\bf B}^k U(1)_{\rm conn}$ or $B^k U(1)$.)
Over any space $M$, $B^k U(1) = K({\mathbb Z},k+1)$ bundles are classified by
homotopy classes of maps into a next higher Eilenberg-Maclane space,
{\it i.e.} homotopy classes of maps
\begin{displaymath}
M \: \longrightarrow \: K({\mathbb Z},k+2),
\end{displaymath}
which are computed by $H^{k+2}(M,{\mathbb Z})$.
Thus, we see that extensions of the form above are classified topologically
by $H^{k+2}(G,{\mathbb Z})$.

Now, in principle we are not yet done.  We need to classify extensions with
higher group structures, not just topological bundles.  For 2-groups of the form
\begin{displaymath}
1 \: \longrightarrow \: {\bf B} U(1) \: \longrightarrow \:
\tilde{G}_1 \: \longrightarrow \: G \: \longrightarrow \: 1,
\end{displaymath}
this second step was performed in \cite{sp}[theorem 99], 
\cite{urs-thesis}[theorem 5.1.29].
An analogous analysis for more general cases is beyond the scope of this
paper, see instead \cite{urs-thesis}[theorem 4.4.36].
See also \cite{bartels1} for a discussion of bundles of 2-groups,
\cite{nss1,nss2} for a discussion of more general bundles of higher groups,
and \cite{kw}[section 3] for a discussion of the relationship between
2-groups and 1-gerbes.

Briefly, the result of the second step is that for $G$ a compact Lie group
of dimension greater than zero, extensions of $G$ by any flavor of
${\bf B}^kU(1)$ are classified by $H^{k+2}(G,{\mathbb Z})$, just as
topological class of bundles.  However, for $G$ finite, there is
additional\footnote{
If $G$ is finite, then for $k \geq 0$, $H^{k+2}(G,{\mathbb Z})$ vanishes,
so the topological bundles are all trivial.  However, one can still have
a nontrivial higher group extension, as this encodes additional structure
on top of the topological bundle.
} information in the second step, and classes of extensions with
higher group structures are classified by group cohomology
$H^{k+2}(G,U(1))$ (with trivial action on the coefficients).

As a sanity check, note in the special case that $k=1$,
for $G$ of dimension greater than zero,
the extensions are classified by $H^3(G,{\mathbb Z})$, as 
consistent with WZW models and Chern-Simons theories, for example.
For $k=1$, if $G$ is finite, the extensions are classified by
$H^3(G,U(1))$, consistent with three-dimensional Dijkgraaf-Witten theory.

\end{document}